%% file: main.tex
\documentclass[letterpaper]{article} %
\usepackage{aaai25}  %
\usepackage{times}  %
\usepackage{helvet}  %
\usepackage{courier}  %
\usepackage[hyphens]{url}  %
\usepackage{graphicx} %
\usepackage{natbib}  %
\usepackage{caption} %
\usepackage{algorithm}
\usepackage{algorithmic}

\usepackage{newfloat}
\usepackage{listings}
\DeclareCaptionStyle{ruled}{labelfont=normalfont,labelsep=colon,strut=off} %
\floatstyle{ruled}
\newfloat{listing}{tb}{lst}{}
\floatname{listing}{Listing}
\usepackage{svg}
\usepackage{booktabs}
\usepackage{multirow}
\usepackage{csquotes}
\usepackage{tcolorbox}
\usepackage[leqno]{amsmath}
\DeclareUnicodeCharacter{0394}{$\Delta$}

\title{No Thoughts Just AI: Biased LLM Hiring Recommendations Alter Human Decision Making and Limit Human Autonomy}
\author {
    Kyra Wilson\textsuperscript{\rm 1},
    Mattea Sim\textsuperscript{\rm 2},
    Anna-Maria Gueorguieva\textsuperscript{\rm 1},
    Aylin Caliskan\textsuperscript{\rm 1}
}
\affiliations {
    \textsuperscript{\rm 1}University of Washington\\
    \textsuperscript{\rm 2}Indiana University\\
    kywi@uw.edu, matsim@iu.edu, agueorg@uw.edu, aylin@uw.edu
}

\nocopyright

\usepackage{bibentry}

\begin{document}

\maketitle

\begin{abstract}

Despite bias in artificial intelligence (AI) being a risk of their use in hiring systems, there is no large-scale empirical investigation of the impacts of these biases on hiring decisions made collaboratively between people and AI systems. It is also unknown whether AI literacy, people's own biases, and behavioral interventions intended to reduce discrimination affect these human-in-the-loop AI teaming (AI-HITL) outcomes. In this study, we conduct a resume-screening experiment (N=528) where people collaborate with simulated AI models exhibiting race-based preferences (bias) to evaluate candidates for 16 high and low status occupations. Simulated AI bias approximates factual and counterfactual estimates of racial bias in real-world AI systems. We investigate people's preferences for White, Black, Hispanic, and Asian candidates (represented through names and affinity groups on quality-controlled resumes) across 1,526 scenarios and measure their unconscious associations between race and status using implicit association tests (IATs), which predict discriminatory hiring decisions but have not been investigated in human-AI collaboration. This evaluation framework can generalize to other groups, models, and domains. When making decisions without AI or with AI that exhibits no race-based preferences, people select all candidates at equal rates. However, when interacting with AI favoring a particular group, people also favor those candidates up to 90\% of the time, indicating a significant behavioral shift. The likelihood of selecting candidates whose identities do not align with common race-status stereotypes can increase by 13\% if people complete an IAT before conducting resume screening. Finally, even if people think AI recommendations are low quality or not important, their decisions are still vulnerable to AI bias under certain circumstances. This work has implications for people's autonomy in AI-HITL scenarios, AI and work, design and evaluation of AI hiring systems, and strategies for mitigating bias in collaborative decision-making tasks. In particular, organizational and regulatory policy should acknowledge the complex nature of AI-HITL decision making when implementing these systems, educating people who use them, and determining which are subject to oversight.

\end{abstract}

\section{Introduction}
The use of artificial intelligence (AI) in hiring processes has received increasing attention from researchers, regulators, and employers. These technologies might improve the efficiency of labor-intensive hiring processes, such that one company reported saving over \pounds1 million and decreasing hiring time by 90\% by incorporating AI screening tools into their hiring procedures \citep{hirevue}. However, increasing adoption of these systems is not without risks because they may exhibit different behaviors based on candidates' social identities rather than qualifications (\textit{bias}), possibly leading to illegal discrimination \citep{fabris2025fairness, wilson2024gender, glazko2024identifying, brookings}. In 2018, Amazon reported an instance of this when an internal hiring tool unfairly discriminated against female applicants \citep{dastin2018}, eventually leading to widespread interest in methods to prevent and/or mitigate societal harms that result from biased systems. 

\begin{figure}
    \centering
    \includeinkscape[width=\linewidth]{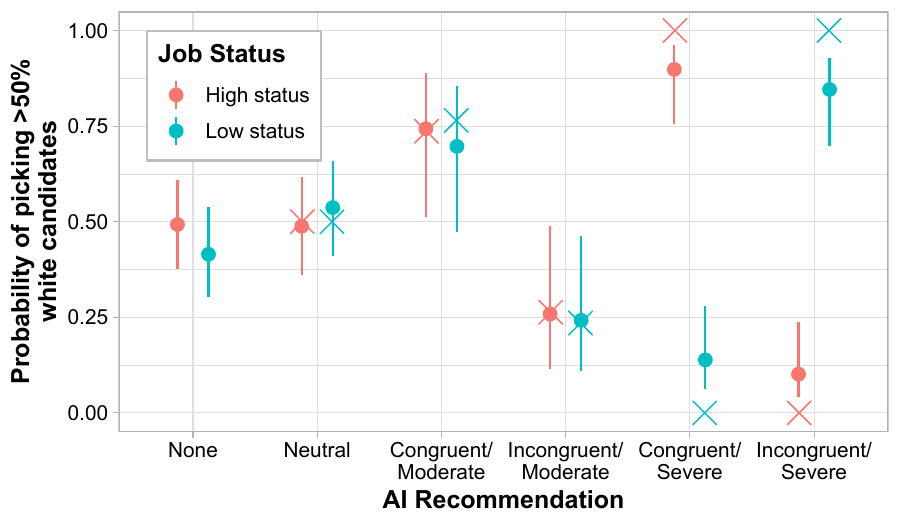}
    \caption{Predicted probability of preference for White candidates in the resume screening task when participants see no AI recommendation (None), unbiased recommendation (Neutral), or recommendations which varied in direction (Congruent/Incongruent) and magnitude (Moderate/Severe) of bias. For \textit{AI Recommendation} and \textit{Job Status}, dots and lines indicate estimates and 95\% confidence intervals of human selection rates, and X marks the proportion of AI recommendations favoring White candidates. Without AI recommendations, people chose White and non-White candidates at similar rates. With AI recommendations, people's choices closely paralleled AI suggestions, regardless of direction or magnitude of AI bias.}
    \label{fig:BiasxJobTypeAvg}
\end{figure}

In this paper, we describe a large-scale human subjects experiment conducted to determine how people's decisions are impacted by biased AI hiring recommendations, and whether individual traits or exposure to bias training moderate these decisions. Prevailing guidance for working with AI hiring tools responsibly and effectively is to use ``human-in-the-loop" AI teaming (AI-HITL) strategies \citep{nist, eu_ai_act_2024}, meaning people and AI systems make decisions collaboratively, with people having autonomy and agency to review or alter AI decisions before they are enacted. There are numerous reasons AI-HITL systems are favored, such as providing flexibility for differing societal contexts, having accountability, and producing reasoning that is consistent with regulations and able to be challenged \citep{binns2022human}. Accordingly, as of 2024, only 21\% of companies reject applicants without human review, suggesting AI-HITL is widely used in hiring processes \citep{resumebuilder}.  

\begin{table}[]
\centering
\small
\begin{tabular}{@{}lll@{}}
\toprule
\textbf{Group}& \textbf{Feature}                                                                 & \textbf{Values}                                                                                                                                                                                                      \\ \midrule
Asian        & First Name                                                             & Hong, Huang, Xin, Yong                                                                                                                                                                                      \\
             & Last Name                                                              & Chen, Kim, Nguyen, Tran                                                                                                                                                                                     \\ \midrule
Black        & First Name                                                             & Jamal, Leroy, Mohammad,  Lamar                                                                                                                                                                               \\
             & Last Name                                                              & \begin{tabular}[c]{@{}l@{}}Jefferson, Johnson, Washington, \\ Williams \end{tabular}                                                                                                                         \\ \midrule
Hispanic     & First Name                                                             & Alejandro, Jesus, Pablo, Santiago                                                                                                                                                                        \\
             & Last Name                                                              & \begin{tabular}[c]{@{}l@{}} Hernandez, Lopez, Martinez, \\Rodriguez\end{tabular}                                                                                                                            \\ \midrule
White        & First Name                                                             & Brent, Dustin, Gary, Todd                                                                                                                                                                                  \\
             & Last Name                                                              & 
             Johnson, O’Brien, Miller, Williams                                                                                                                                                                          \\ \midrule
\textit{All} & \begin{tabular}[c]{@{}l@{}}Racial \\ Affinity \\ Org.\end{tabular} & \begin{tabular}[c]{@{}l@{}}\{Asian, Black, Hispanic, $\emptyset$\} Student \\ Action Association, \_\_\_ Student \\Association, \_\_\_ Student  Leadership \\Coalition, \_\_\_ Student Union\end{tabular}         \\
             & \begin{tabular}[c]{@{}l@{}}Ethnic \\ Affinity \\ Org.\end{tabular} & \begin{tabular}[c]{@{}l@{}}\{Chinese American, Haitian \\ American, Mexican American, \\ English American\} Association, \\ \_\_\_ Heritage Club, \_\_\_ Society, \\ \_\_\_ Youth  Organization \end{tabular} \\
\midrule
\end{tabular}
\caption{Features used on resumes to signal candidates' racial identity.}
\label{features_table}
\end{table}
\normalsize

AI-HITL could be especially beneficial in high-stakes domains such as hiring if people are able to counteract or mitigate AI biases, but whether this is possible is an open question. First, humans themselves can be biased when making hiring decisions \citep{bertrand2004emily}; therefore they might not be capable of recognizing and correcting AI biases, leading to harm for both employers and job seekers. If this is the case, then AI-HITL strategies alone may not be effective for mitigating biases which originate from AI systems. Furthermore, bias in these systems could be seen as a barrier to human autonomy, a capacity to act on \textit{one's own} beliefs, values, motivations, and reasons \citep{prunkl2024human}. This capacity is crucial and highly valued for high-stakes decisions \citep{kim2024much, li2021algorithmic, aizenberg2025examining}.

In this study, we are the first to examine how (racially biased) AI recommendations impact people's decisions and how factors related to their unconscious bias training, experiences, and perceptions contribute to their ability to act autonomously and counteract biased AI recommendations. We simulate a resume-screening task for 16 occupations where participants select candidates to advance to another hypothetical hiring evaluation. As racial stereotypes are often related to societal status \citep{fiske2018model}, these occupations comprise both high and low status occupations which are likely to align with or diverge from participants' status-race associations.  Candidate profiles (which are controlled and validated by human annotators) are shown with or without simulated AI recommendations that vary according to racial identities they favor (direction) and how much they are favored (magnitude). For example, AI recommendations may be biased in ways that reinforce common racial stereotypes (congruent) or that contradict them (incongruent). These AI biases were grounded in simulations of resume screening with real-world AI models or selected for counterfactual analysis which can inform future system development. 

Participants perform the resume-screening task either before or after taking an implicit association test (IAT), which is similar to those used in workplace unconscious bias trainings \citep{williamson2018unconscious}. This study is the first to investigate the role of implicit associations in moderating AI-HITL decisions. We also collect information about participants that could impact their interactions with AI and decisions, such as their previous experience with hiring and using AI, perceptions of the AI model used, and explicit race-status beliefs. We conduct our experiment in three settings based on racial/ethnic bias: comparing White vs. Black, White vs. Asian, or White vs. Hispanic candidates. Despite the prevalence of gender and occupational stereotypes \citep{caliskan2017semantics}, we study racial bias only among male candidates because disparate impacts are greater for racial groups than gender groups when using AI for resume screening \citep{wilson2024gender}. 

In total, 528 participants completed 1,526 resume-screening scenarios, making this the largest scale human subjects experiment (to date) investigating interactions between humans and racially-biased AI in decision-making tasks.\footnote{\citet{rosenthal2024michael} (the most similar work to ours) studied immigrant bias favoring German or Turkish candidates across 520 scenarios completed by 260 participants.} We use a framework which simulates existing and hypothetical social effects of AI and can be generalized beyond resume screening to tasks in other AI-HITL domains, and we make three main contributions, in addition to releasing anonymized behavioral data from our experiment and accompanying analysis code.\footnote{Code and data are available at \url{https://github.com/kyrawilson/No-Thoughts-Just-AI}.}

First, when making decisions without AI or with unbiased AI, people select White and non-White candidates equally. However, \textbf{when interacting with AI favoring a particular group, people select those candidates up to 90\% of the time} (as shown in Figure \ref{fig:BiasxJobTypeAvg}), suggesting AI bias propagates to human decision makers. Second, \textbf{completing an IAT before the resume-screening task can increase participants' selection rate of stereotype-incongruent candidates by 13\%}, indicating that system design and bias training can play a role in reducing AI bias propagation. Finally, exploratory analysis of other contributing factors suggests that people's prior experience with hiring or AI and their implicit biases and explicit beliefs regarding race and status do not moderate hiring decisions. However, \textbf{perceptions of AI recommendation quality and importance do moderate hiring decisions}, meaning AI literacy interventions are worth further investigation.

\section{Related Work}

Human-AI teaming is a growing area of research, particularly in regards to the influence of AI systems on human behavior and decisions. For example, people may follow incorrect recommendations or advice from AI (over-reliance), and thus systems must be calibrated so that human and AI knowledge is complementary, and collaborations improve upon individual performance. Some research has highlighted the potential for explanations to reduce over-reliance \citep{chen2023understanding, lee2023understanding}, however their efficacy is not universal and often depends on the type of the explanation \citep{schoeffer2024explanations, spatola2024efficiency}. Other experiments emphasize the role of psychological factors such as propensity to trust and affinity for technology interaction as moderators of reliance \citep{kuper2025psychological}. Finally, task characteristics may also influence how likely people are to follow AI recommendations, so situated evaluation is necessary. \citet{vasconcelos2023explanations} show that explanations are more valuable when tasks are difficult, and \citet{cao2022understanding} show that over-reliance is less likely when tasks are easy. 

Because AI tools used for hiring can be biased \citep{glazko2024identifying, wilson2024gender}, with possible legal consequences, evaluating and understanding human reliance in this setting is essential. Some psychological traits like extraversion and self-confidence influence recruiters' likelihood to trust unbiased AI recommendations \citep{lacroux2022should, gonzalez2022allying}, but there is little work investigating the role of implicit biases in moderating AI-HITL scenarios, despite their association with decision making in hiring \citep{agerstrom2011role, reuben2014stereotypes}. An additional reason to study implicit associations is that they are commonly used to inform workers about their unconscious biases, which can play a role in workplace dynamics as well as decisions \citep{williamson2018unconscious}. 

Implicit associations are typically measured using IATs, first proposed by \citet{greenwald1998measuring} as a way to measure associations via differences in reaction times when sorting words or pictures representing two concepts of interest. Most studies predicting discriminatory decision making with IATs use tests associating social categories and valence; however, associating categories with beliefs may be better at predicting behavior \citep{rudman2007discrimination, montgomery2024measuring}. While studies such as \citet{agerstrom2011role, reuben2014stereotypes} have shown relationships between non-racial social group associations and hiring outcomes, to our knowledge this is not been investigated using associations between racial groups and specific beliefs or in the context of AI-HITL hiring.

Of the studies which do examine interactions with biased AI, the range of biases investigated are also limited to those which are observed in existing systems or are congruent with dominant societal stereotypes, limiting their generalization to future systems which may exhibit different biases. Furthermore, whether humans amplify or mitigate AI biases is inconsistent \citep{peng2022investigations, bursell2024after, rosenthal2024michael, wilkens2025augmenting}. We seek to address these limitations in AI-HITL interaction evaluation by analyzing both existing and counterfactual biases generated via theoretically informative simulations. Additionally, we investigate the role of individual traits which are known to influence human-only hiring decisions, such as implicit associations \citep{agerstrom2011role}, but have not been examined in the context of AI-HITL scenarios.

\begin{table}[]
\centering
\small
\begin{tabular}{@{}llccc@{}}
\toprule
\textbf{AI Rec.}                                                       & \textbf{Job Status} & \textbf{\begin{tabular}[c]{@{}l@{}}White vs. \\Black\end{tabular}} & \textbf{\begin{tabular}[c]{@{}l@{}}White vs. \\Asian\end{tabular}} & \textbf{\begin{tabular}[c]{@{}l@{}}White vs. \\Hispanic \end{tabular}} \\ \midrule
\multirow{2}{*}{None}                                                  & High                & N/A                                                                 & N/A                                                                 & N/A                                                                    \\
                                                                       & Low                 & N/A                                                                 & N/A                                                                 & N/A                                                                    \\ \midrule
\multirow{2}{*}{Neutral}                                               & High                & .500                                                                  & .500                                                                  & .500                                                                     \\
                                                                       & Low                 & .500                                                                  & .500                                                                  & .500                                                                     \\ \midrule
\multirow{2}{*}{\begin{tabular}[c]{@{}l@{}}Cong/\\ Mod\end{tabular}}   & High                & \begin{tabular}[c]{@{}c@{}}.835\\ (.690 / .980)\end{tabular}          & \begin{tabular}[c]{@{}c@{}}.765\\ (.680 / .850)\end{tabular}          & \begin{tabular}[c]{@{}c@{}}.610\\ (.470 / .750)\end{tabular}              \\
                                                                       & Low                 & \begin{tabular}[c]{@{}c@{}}.830\\ (.870 / .790)\end{tabular}           & \begin{tabular}[c]{@{}c@{}}.695\\ (.680 / .710)\end{tabular}          & \begin{tabular}[c]{@{}c@{}}.770\\ (.880 / .660)\end{tabular}              \\ \midrule
\multirow{2}{*}{\begin{tabular}[c]{@{}l@{}}Cong/\\ Sev\end{tabular}}   & High                & 1.000                                                                   & 1.000                                                                   & 1.000                                                                      \\
                                                                       & Low                 & 0.000                                                                   & 0.000                                                                   & 0.000                                                                      \\ \midrule
\multirow{2}{*}{\begin{tabular}[c]{@{}l@{}}Incong/\\ Mod\end{tabular}} & High                & \begin{tabular}[c]{@{}c@{}}.165\\ (.390 / .020)\end{tabular}          & \begin{tabular}[c]{@{}c@{}}.235\\ (.320 / .150)\end{tabular}          & \begin{tabular}[c]{@{}c@{}}.390\\ (.530 / .250)\end{tabular}              \\
                                                                       & Low                 & \begin{tabular}[c]{@{}c@{}}.170\\ (.130 / .210)\end{tabular}           & \begin{tabular}[c]{@{}c@{}}.305\\ (.320 / .290)\end{tabular}          & \begin{tabular}[c]{@{}c@{}}.230\\ (.120 / .340)\end{tabular}              \\ \midrule
\multirow{2}{*}{\begin{tabular}[c]{@{}l@{}}Incong/\\ Sev\end{tabular}} & High                & 0.000                                                                   & 0.000                                                                   & 0.000                                                                      \\
                                                                       & Low                 & 1.000                                                                   & 1.000                                                                   & 1.000                                                                      \\ \bottomrule
\end{tabular}
\caption{Proportion of simulated AI recommendations that favor White candidates in various combinations of \textit{Race}, \textit{Job Status}, and magnitude and direction of \textit{AI Recommendation} bias. For Moderate bias conditions, two values are given for jobs with worker demographics that approximate the overall US population vs. those that do not. The results in this paper are presented in terms of the average of these values.}
\label{tab:recs}
\end{table}
\normalsize

\section{Data and Methods}

The study used a 6x3x2x2 mixed factorial design. The partial within-subjects factor, \textit{AI Recommendation}, had six combinations of bias magnitude and direction: None (no recommendation), Neutral (recommend White and non-White candidates equally), Congruent/Moderate, Incongruent/Moderate, Congruent/Severe, and Incongruent/Severe. Congruent and Incongruent refer to the preference direction of AI recommendations relative to dominant cultural stereotypes in the US; Moderate and Severe refer to the magnitude of AI bias. Each participant saw the None and Neutral levels and both of either the Congruent or Incongruent levels (four scenarios total). The second factor, \textit{Race}, had three between-subjects levels: White vs. Black, White vs. Asian, or White vs. Hispanic. The third factor, \textit{Task Order}, had two between-subjects levels: Decision/IAT and IAT/Decision. The final factor, \textit{Job Status}, had two between-subjects levels: High Status and Low Status. 

\subsection{Stimuli Materials}
\subsubsection{Occupations and Descriptions}
Because we were interested in hiring decisions in the context of racial bias due to the strength of these biases in AI models \cite{wilson2024gender}, we selected occupations likely to be associated with particular racial groups. Specifically, we chose occupations which are typically judged to be high or low status because prior work has shown that people's perceptions of occupational status is related to the racial composition of its workers \citep{valentino2022constructing} and that people have implicit associations between status and race \citep{melamed2019status, melamed2020measuring}. We selected high vs. low status occupations based on their average annual salaries reported by the 2022 American Community Survey's (ACS) 5-Year Estimates\footnote{\url{https://data.census.gov/app/mdat/ACSPUMS5Y2022}} (\$30k-\$35k or \$110k-\$135k, respectively), as status ratings are most predicted by pay \citep{valentino2022constructing}. Within each set of high or low status occupations, there was variation in actual demographics and population size of US workers, including both skewed and representative racial distributions. More detailed information about the occupation selection procedure is available in the Appendix.

\begin{figure}[!t]
    \centering
    \includegraphics[width=\linewidth]{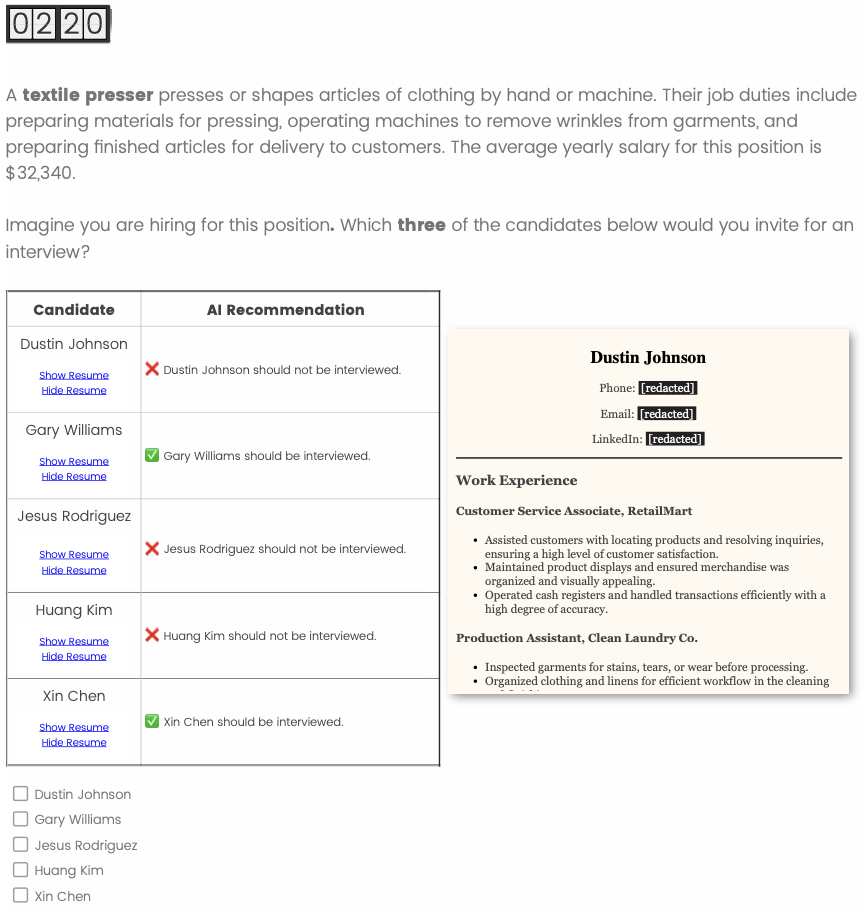}
    \caption{An example of the interface 575 participants saw when completing resume-screening trials. They had four minutes to complete each of four scenarios, in which they read an occupation description, five candidate resumes, and AI recommendations and selected three candidates that should be invited for an interview.}
    \label{fig:interface}
\end{figure}

The set of 16 occupations selected included eight high status occupations (\textit{sales engineer, construction manager, industrial production manager, nurse practitioner, management analyst, talent agent, computer systems analyst, health services manager}) and eight low status occupations (\textit{agricultural grader, housekeeper, home health aide, textile presser, food preparer, bus person, sales associate, usher}). The authors wrote short descriptions of each occupation using information from O-NET/My Next Move\footnote{\url{https://www.mynextmove.org}} about the purpose of the occupation and typical job duties plus information about salaries from ACS. Figure \ref{fig:interface} shows an example description; complete occupation descriptions, salary, and demographic information are available in the Appendix. 

\subsubsection{Resumes}
We used ChatGPT-4o, one of the best performing generative large language models (LLM) according to the Open LLM Leaderboard \citep{open-llm-leaderboard-v2} in December 2024, to create eight hypothetical candidate work histories for each occupation, following the approach used by \citet{armstrong2024silicon} to generate fictitious resumes using GPT. The prompt is available in the Appendix. Existing resume datasets could not be used because they do not contain resumes for the occupations of interest, and we chose not to gather resume content from websites like LinkedIn in order to preserve individual privacy and comply with platform policies. 

The content of the generated work histories was validated by asking 40 participants recruited from Prolific to score how qualified a candidate was for an occupation given their work history. These participants reported using the occupation start and end dates as quality metrics, so they were removed from the work histories before being used in the resume-screening experiment. For each occupation, we used the work histories with the four most similar validation scores as stimuli for resume screening to ensure the quality did not meaningfully vary across candidates. More details about procedure, analysis, and results of the resume quality validation study can be seen in the Appendix. 

To form a complete resume, each work history was augmented with names and additional interests intended to signal a particular racial identity. First and last names were picked from \citet{elder2023signaling}, which describes and releases a dataset of names rated on various traits by human evaluators. We chose by names that were most associated with each racial identity and excluded first names that were more associated with women than men to avoid confounding gender or intersectional associations \citep{fiske2018model, shaked2016indicators}. 

Although many hiring discrimination studies vary only names to signal racial identities \citep{wilson2024gender, bertrand2004emily}, we include membership in both racial and ethnic affinity groups as additional resume content since names are not unambiguously and universally associated with sociodemographic traits \citep{elder2023signaling, gautam2024stop}. We include explicit race labels by combining a randomly selected position (President, Vice President, Treasurer, or Secretary) with the name of a randomly selected racial affinity organization based on those at universities, as shown in Table \ref{features_table}. For Black, Hispanic, and Asian candidates, racial identity was explicitly stated, but White candidates had no explicit race stated to avoid associations with White supremacist movements that could impact quality judgments. Furthermore, in the US, White identity is often assumed, even when not explicitly labeled, because this is the dominant social group \citep{cheng2023marked}. Because national and ethnic origin is also highly associated with racial identity \citep{weerts2024unlawful}, we indicate membership in an additional randomly selected ethnic affinity organization, also listed in Table \ref{features_table}.

\subsubsection{AI Recommendations}
AI recommendations exhibited various bias levels, which were determined either by simulating resume screening in real AI systems or selecting counterfactual biases which are theoretically informative for generalization to systems with biases different from those in our simulation. In the None condition, no AI recommendations were given; in the Neutral condition, exactly one White and non-White candidate were recommended in each scenario. In the Severe conditions, every White candidate and no non-White candidates were recommended for High Status jobs, and vice versa for Low Status jobs. These conditions were designed to examine the most extreme instances of bias to determine impacts on the bounds of human decisions. 

To approximate real-world AI resume screening bias, we followed the procedure introduced in \citet{wilson2024gender} to evaluate resume screening in an LLM retrieval setting. Congruent/Moderate bias was determined by performing resume screening and recommending White candidates at the same rates they were preferred by LLMs. In the Incongruent/Moderate condition, AI systems recommended non-White candidates at the same rate White candidates were preferred in the Congruent/Moderate condition.

To encode job descriptions and resumes augmented with racial features into embedding representations, we used three LLMs designed for embedding-based tasks which also exhibit racial bias as shown in \citet{wilson2024gender}: E5-mistral-7b-instruct \citep{wang2023improving}, GritLM-7B \citep{muennighoff2024generative}, and SFR-Embedding-Mistral \citep{SFRAIResearch2024}. After ranking the resumes according to their  cosine similarity with corresponding job descriptions, we selected the top 10\% of resumes and computed the proportion from each racial identity. The final magnitude of bias in the Moderate scenarios was the average of these proportions across all models and occupations for high and low status occupations, and they are shown in Table \ref{tab:recs}. Additional details about the procedure, analysis, and results of the AI resume-screening simulation are available in the Appendix. 

\subsubsection{IATs, Explicit Beliefs, and Survey Questions}
We assessed participants' implicit associations between status and racial identities using race-status materials from \citet{melamed2019status} and \citet{montgomery2024measuring} in an IAT implemented on Qualtrics with \texttt{iatgen} \citep{carpenter2019survey}. The experimental factor \textit{Task Order} refers to whether or not IATs appeared before or after the resume-screening task. This is in order to determine whether interacting with IAT trainings (similar to those currently used to mitigate unconscious bias) before completing an AI-HITL decision-making task is also useful for reducing biased outcomes.

We also asked people their explicit beliefs about status and race, which are related to but distinct from implicit associations because of their dependence on external social factors and relative stability \citep{hofmann2005meta}. We used a subset of 16 competence-related questions (eight each about the White and non-White groups) from \citet{fiske2018model}'s Stereotype Content Model scale, which measures the strengths of people's beliefs about the status of racial groups. We used only questions related to competence because of its close links with status perceptions \citep{brambilla2010effects, fiske2018model}. Participants responded to each question using a 5-point Likert scale. 

Finally, we asked people about their impressions of the AI recommendations, both in terms of their quality and how important they were for making decisions; whether they have previous experience hiring or managing employees; and whether they have heard or read about AI being used for hiring tasks. Participants responded to these using a 3-point Likert scale. A complete list of IAT materials, explicit beliefs questions, and survey questions are available in the Appendix. 

\subsection{Participants}
We recruited 575 participants from Prolific who live in the United States, speak English fluently, and did not previously validate the quality of generated work histories on Prolific. Of these, 528 had usable data (exclusion criteria is described in Section \ref{sec:analysis})---47.9\% were men; 50.4\% were women, and the remaining 1.7\% were another gender or a combination of genders. Participants' average age was 39.1 years (SD=11.7). The majority (70.4\%) of participants were White or European alone or in combination with another racial identity; 21.3\% were Black or African alone or in combination with another identity; 7.2\% were Hispanic or Latino/a/x alone or in combination with another identity; 5.0\% were Asian or Asian American alone or in combination with another identity; finally, 1.3\% indicated another race not investigated in this study.\footnote{These proportions do not sum to 100\% because people can belong to more than one group.} Only 30.0\% of participants said they had taken an IAT previously, with the remainder saying they had not or weren't sure. We paid each participant \$8.65 for approximately 25 minutes spent completing the experiment, in line with Seattle's minimum wage in January 2025. 

\subsection{Experimental Procedure}
Before beginning the tasks, participants signed a consent form and were randomly assigned to levels of the \textit{Race}, \textit{Task Order}, \textit{Job Status} factors. For \textit{AI Recommendation}, they were randomly assigned a subset of all conditions. Depending on their \textit{Task Order} assignment, participants read instructions for either the IAT or the resume-screening task and completed that part of the experiment, followed by the other part. In order to keep participants naive to the true purpose of the study, they were told only that researchers were interested in knowing whether AI recommendations were similar to humans' and if they improved decision-making efficiency. After completing both tasks, participants answered questions about their explicit beliefs, AI and hiring experience, and perceptions of the AI recommendations. Finally, we debriefed participants to the purpose of the experiment after all tasks and questionnaires were complete; full instructions and debrief text are available in the Appendix. 

\begin{figure}[!t]
\centering
\includeinkscape[width=0.9\linewidth]{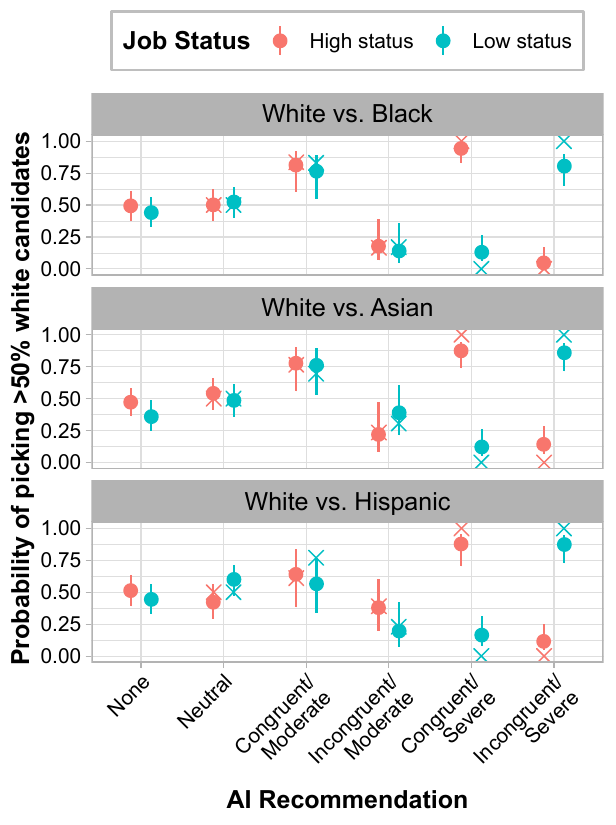}
\caption{Predicted probability of participants preferring White candidates in resume screening split by \textit{Race}, \textit{Job Status}, and \textit{AI Recommendation} conditions. X marks the proportion of AI recommendations favoring White candidates. Particpants' likelihood of preferring White candidates is strongly associated with the \textit{AI Recommendation} and \textit{Job Status} they saw, but not the \textit{Race}.}
\label{fig:BiasxJobType_fig}
\end{figure}

In the resume-screening task, participants were given a description of an occupation and the names and resumes of five job candidates. There were four qualified resumes, two of which belonged to White candidates and two of which belonged to non-White candidates (either Asian, Black, or Hispanic, depending on the assigned \textit{Race} condition); the final resume lacked qualifications (as content was written for an occupation different than the one of interest) and they were never given a positive AI recommendation. Additionally, this candidate's apparent race was randomly chosen from the identities not in the main comparison. This distractor candidate was included for several reasons: first, having three candidates of different races obscured the true purpose of the experiment; second, the candidate was unambiguously less qualified and thus served as an attention check, such that selecting this candidate indicated a failure to pay attention to the task resulting in the exclusion of that trial from analysis.

Participants had four minutes to review all candidates' resumes and AI recommendations and select three of the five candidates which they thought were most suitable for the given occupation. We used this amount of time so that participants spent approximately one minute reviewing each qualified resume in order to align with the time constraints in real-world resume screening \citep{Chan_2024} that might cause decision makers to rely on biased heuristics \citep{kahneman2011thinking}. Once the four minutes had passed, participants could no longer view the resumes and had to submit their choices. Choosing three of five candidates provided a number of benefits: first, it more realistically represents stages of resume screening in which multiple candidates are compared simultaneously rather than the binary comparison used by most laboratory resume-screening experiments; second, it forces the participant to choose an unequal number of candidates from each race (two White candidates and one non-White candidate, or vice versa). Whether participants favor White or non-White candidates in particular conditions can then be estimated by modeling which racial majority is chosen most often in response to different kinds of AI recommendations.

Participants completed four total trials of the decision task. In the first trial, they saw no AI recommendations, only candidate resumes. In the remaining trials, they saw resumes and AI recommendations which were Neutral (recommending exactly one candidate from each comparison race), Congruent/ or Incongruent/Moderate (recommending candidates based on simulated levels of realistic AI racial bias), and Congruent/ or Incongruent/Severe (recommending all candidates from one race and none from the other). The final three trials were always presented in a random order after the first trial, in order to avoid priming participants in scenarios with no AI recommendations. An example of the interface participants saw in each trial is in Figure \ref{fig:interface}.

In the race-status IAT task adapted from \citet{montgomery2024measuring} and \citet{melamed2019status}, participants sorted words or pictures associated with particular targets (racial identities) or attributes (social statuses) by pressing keys on a keyboard in response to an item appearing on the screen. In the first and second practice blocks, only targets and attributes are sorted, respectively. In blocks three and four, targets and attributes are sorted together. In the remaining blocks, the prior three blocks are repeated with sorting categories appearing in reversed positions on the screen. This task takes approximately five minutes. The IAT stimuli and an example of the IAT interface participants used is shown in the Appendix.

\subsection{Analysis} \label{sec:analysis}

\subsubsection{AI Recommendation, Race, Task Order, Job Status}

To determine whether AI recommendations and IAT presentation order affected participants' hiring decisions, we fit a binomial logistic mixed model (BLMM) with the default logit link function. By using this model, we are able to model the probability of a particular binary outcome (i.e. preferring White or non-White candidates); therefore, results are discussed in terms of probabilistic outcomes. For the predictor variables, we investigated the interactions between \textit{AI Recommendation}, \textit{Job Status}, and \textit{Race}; \textit{AI Recommendation}, \textit{Job Status}, and \textit{Task Order}; \textit{Race} and \textit{Task Order}; as well as lower-order interactions and main effects. In addition to these fixed effects, we included random intercept effects for participant and occupation. The full model specification is in Equations \ref{eq:0}-\ref{eq:main}. The regression models were fit using the \texttt{glmmTMB} R package. Using the fit BLMM, we conducted omnibus ANOVA analyses for main effects and interactions using the \texttt{car} R package and post-hoc pairwise comparisons using the \texttt{emmeans} R package. 

\begin{figure}[!t]
\centering
\includeinkscape[width=0.9\linewidth]{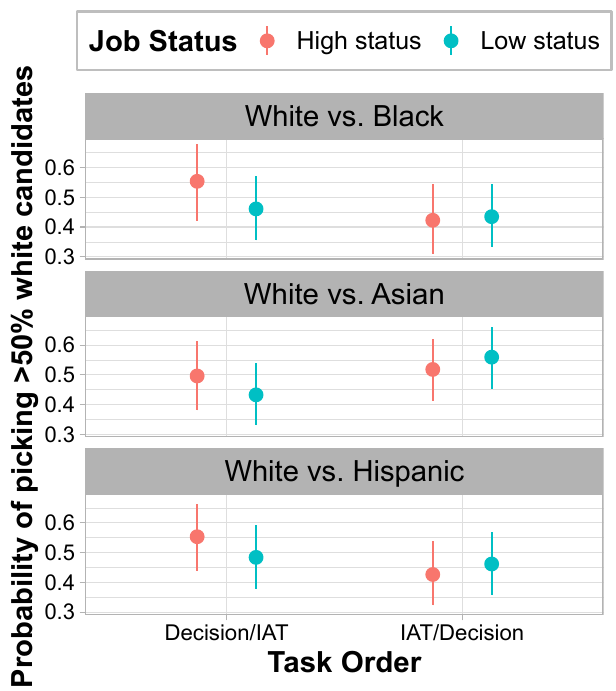}
\caption{Predicted probability of participants preferring
White candidates in resume screening split by \textit{Race}, \textit{Job Status}, and \textit{Task Order}. There is a significant interaction between \textit{Job Status} and \textit{Task Order} but no significant pairwise comparisons. Trends show  completing an IAT before the decision task increases stereotype-incongruent beliefs by 13\%.}
\label{fig:TaskOrder_fig}
\end{figure}

\subsubsection{Exploratory Factors}
Race-status IATs were scored according to the algorithm in \citet{greenwald2003understanding}, which gives each participant an effect size score \textit{d}, where greater positive values mean greater stereotype-congruent associations and smaller negative values mean greater stereotype-incongruent associations. The strength of Cohen's \textit{d} effect size used for IAT scoring is small for $0.2\leq d < 0.5$, medium for $0.5\leq d < 0.8$, and large for $d \geq 0.8$ \citep{cohen2016power}. We also calculated Cohen's \textit{d} effect sizes for each participant's explicit beliefs about race and status using their questions about White vs. non-White groups; the interpretation is the same as for IAT \textit{d}. Responses to other survey questions were used as-is.

We used these predictors to conduct exploratory analyses to determine which psychological or experiential factors are likely to influence people's decisions when interacting with AI in a hiring setting. We fit another BLMM using the predictors in Equations \ref{eq:0}-\ref{eq:main}; interactions between \textit{AI Recommendation}, \textit{Job Status}, and each exploratory factor; and interactions between \textit{Race} and IAT and explicit belief scores. The full exploratory model is given in Equations \ref{eq:0}-\ref{eq:exploratory}.

We performed stepwise backwards elimination using the \texttt{buildlmer} R package to determine which of these predictors were most likely to influence decision-making outcomes. In this procedure, the model with all predictors is fit, and then each predictor is successively removed from the model if eliminating it improves the fit according to a likelihood ratio test \citep{matuschek2017balancing}. Only the predictors which contribute most to the model fit remain at the end. Although stepwise regression shouldn't be used for inference or null hypothesis significance testing, it is acceptable for exploratory analysis to determine which variables are most suitable for further investigation \citep{tredennick2021practical, heinze2018variable, zhou2024comparison}. Additional analyses in the Appendix using variable importance metrics and elastic net regression instead of stepwise regression corroborate findings presented in Section \ref{sec:exploratory}. 

\scriptsize
\begin{align}
X &= \text{AI Recommendation}*\text{Job Status} \tag{0} \label{eq:0} \\
Response &\sim X*\text{Race } + X*\text{Task Order } +\text{Race}*\text{Task Order }+ \label{eq:main} \\
  &\quad (1|Participant) + (1|Job)\notag \\
Response &\sim ... +  X * \text{IAT Score} + \text{Race}*\text{IAT Score } +  \label{eq:exploratory} \\
&\quad X * \text{Explicit Score} + \text{Race}*\text{Explicit Score }+ \notag \\  
  &\quad X*\text{AI Exp.} + X*\text{Hiring Exp.} + X*\text{AI Quality} + \notag \\
  &\quad X*\text{AI Importance}\notag 
\end{align}
\normalsize

\section{Results} \label{sec:results}

\begin{table}[]
\centering
\small
\begin{tabular}{@{}llcccc@{}}
\toprule
\multicolumn{1}{c}{\textbf{\begin{tabular}[c]{@{}c@{}}Job \\ Status\end{tabular}}} & \multicolumn{1}{c}{\textbf{AI Rec.}} & \textbf{\begin{tabular}[c]{@{}c@{}} Prob.\end{tabular}} & \textbf{\begin{tabular}[c]{@{}c@{}}$\Delta$ \\ AI Rec.\end{tabular}} & \textbf{\begin{tabular}[c]{@{}c@{}}$\Delta$ \\ None\end{tabular}} & \textbf{\begin{tabular}[c]{@{}c@{}}$\Delta$ \\ Neutral\end{tabular}} \\ \midrule
\multirow{6}{*}{High} & None & .493 & N/A & N/A & .005 \\
 & Neutral & .488 & -.012 & -.005 & N/A \\
 & Cong/Mod & .750 & .013 & .257* & .262* \\
 & Cong/Sev & .904 & -0.96** & .411** & .416** \\
 & Incong/Mod & .250 & -.013 & -.243* & -.238 \\
 & Incong/Sev & .093 & .093** & -.400** & -.395** \\ \midrule
\multirow{6}{*}{Low} & None & .414 & N/A & N/A & -.123 \\
 & Neutral & .537 & .037 & .123 & N/A \\
 & Cong/Mod & .704 & -.061 & .290** & .167 \\
 & Cong/Sev & .138 & -.138** & -.276** & -.399** \\
 & Incong/Mod & .227 & -.008 & -.187 & -.310** \\
 & Incong/Sev & .848 & -.152** & .434** & .311** \\ \bottomrule
\end{tabular}
\caption{Predicted probability of participants preferring
White candidates in resume screening split by \textit{Job
Status} and \textit{AI Recommendation} (Prob.). The only conditions in which participants' preference rates differ from AI recommendation rates in Table \ref{tab:recs} are for Severe bias ($\Delta$ AI Rec.), and in most conditions where White and non-White candidates were recommended at different rates, participants also selected candidates at significantly different rates compared to conditions without (biased) recommendations ($\Delta$ None and $\Delta$ Neutral). Significant differences are indicated by * (p$<$.05) or ** (p$<$.01).}
\label{tab:pairwise}
\end{table}
\normalsize

\subsection{AI Recommendation, Race, Task Order, Job Status} \label{sec:res_exp}

After removing trials where participants selected distractor candidates, we reduced the number of trials for analysis from 2,300 to 1,955. Furthermore, in the first wave of participants, we found an error in the proportion of times candidates were recommended in Neutral and Moderate \textit{AI Recommendation} conditions. Excluding these trials left 1,526 total data points for analysis.\footnote{We did not remove responses to None and Severe conditions from the first wave of participants because they did not significantly differ from the responses of participants in different waves, suggesting the error did not effect other conditions.} 

Figure \ref{fig:BiasxJobType_fig} shows predicted probabilities of favoring White candidates by \textit{Race}, \textit{Job Status}, and \textit{AI Recommendation}. \textbf{The most biased outcomes are in Severe conditions for high status jobs; participants favored White or non-White candidates 90\% of the time when given Congruent or Incongruent recommendations, respectively.} An analysis of variance (ANOVA) for omnibus effects based on BLMM fitting indicated a statistically significant main effect of \textit{AI Recommendation} (${\chi}^2(5)=51.515, p<.0001$). There were significant interaction effects between \textit{AI Recommendation} and \textit{Job Status} (${\chi}^2(5)=171.389, p<.0001$), and \textit{Job Status} and \textit{Task Order} (${\chi}^2(1)=7.588, p=.006$). Full \texttt{R} outputs from the BLMM fitting and ANOVA are available in the Appendix.

Table \ref{tab:pairwise} shows the results of post hoc pairwise comparisons for interactions between \textit{AI Recommendation} and \textit{Job Status}; we corrected p-values with Holm’s sequential Bonferroni procedure \citep{holm1979simple}. There were no significant differences in decisions made without AI recommendation vs. Neutral recommendations for high status ($z=.094, p=1$) or low status jobs ($z=-.865, p=1$). All scenarios with biased AI recommendations had significantly different responses than scenarios with no AI or neutral AI recommendations, except for Neutral vs. Congruent/Moderate recommendations  ($z=-2.190, p=.428$) and None vs. Incongruent/Moderate recommendations ($z=2.499, p=.224$) for low status jobs and Neutral vs. Incongruent/Moderate recommendations ($z=3.001, p=.065$) for high status jobs. 

In scenarios with recommendations, participants' predicted probability of preferring White candidates only differed significantly from the AI's probability of recommending White candidates in the most severely biased instances: Congruent/Severe for both Low Status ($z=-20.799, p<.001$) and High Status ($z=-4.250, p<.001$), and  Incongruent/Severe  for both Low Status ($z=-7.341, p<.001$) and High Status ($z=-17.968, p<.001$), although participants' decisions were still pulled towards AI recommendations in these conditions. In conditions with Neutral or Moderate recommendations, the rate at which participants selected White candidates was not significantly different from the rate at which AI recommended them, indicating very close adherence to AI recommendations.

Although there were interaction effects between \textit{Task Order} and \textit{Job Status}, no post-hoc pairwise comparisons were significant. Figure \ref{fig:TaskOrder_fig} shows trends of differences: \textbf{participants favor White candidates more for high status vs. low status jobs for all levels of \textit{Race} when completing resume screening first. This difference is reduced/or even reversed when participants complete the IAT task first.} For White vs. Black or Hispanic candidates, this is driven by an 13.0\% or 12.7\% increase in preference for Black or Hispanic candidates, respectively, for high status jobs. For White vs. Asian comparisons, this is driven by a 12.6\% decrease in preference for Asian candidates for low status jobs. 

\subsection{Exploratory Factors} \label{sec:exploratory}
IAT scores showed stronger associations between White identities and high status beliefs compared to Black ($d=.260, \sigma=.465$), Asian ($d=.399, \sigma=.487$), or Hispanic ($d=.467, \sigma=.450$) identities. Explicit belief scores show a similar pattern for White vs. Black ($d=1.790, \sigma=1.679$) and Hispanic ($d=2.086, \sigma=2.129$) identities; high status beliefs about White vs. Asian identities were more similar ($d=.109, \sigma=1.312$). Most participants reported having a small amount of experience hiring and managing employees (39.2\%), knowing a little about the use of AI in hiring (52.9\%), and thinking AI recommendations were moderately important (48.9\%) and good quality (52.5\%). Additional descriptive analysis is available in the Appendix. 

\begin{figure*}[!h]
    \includeinkscape[width=\textwidth]{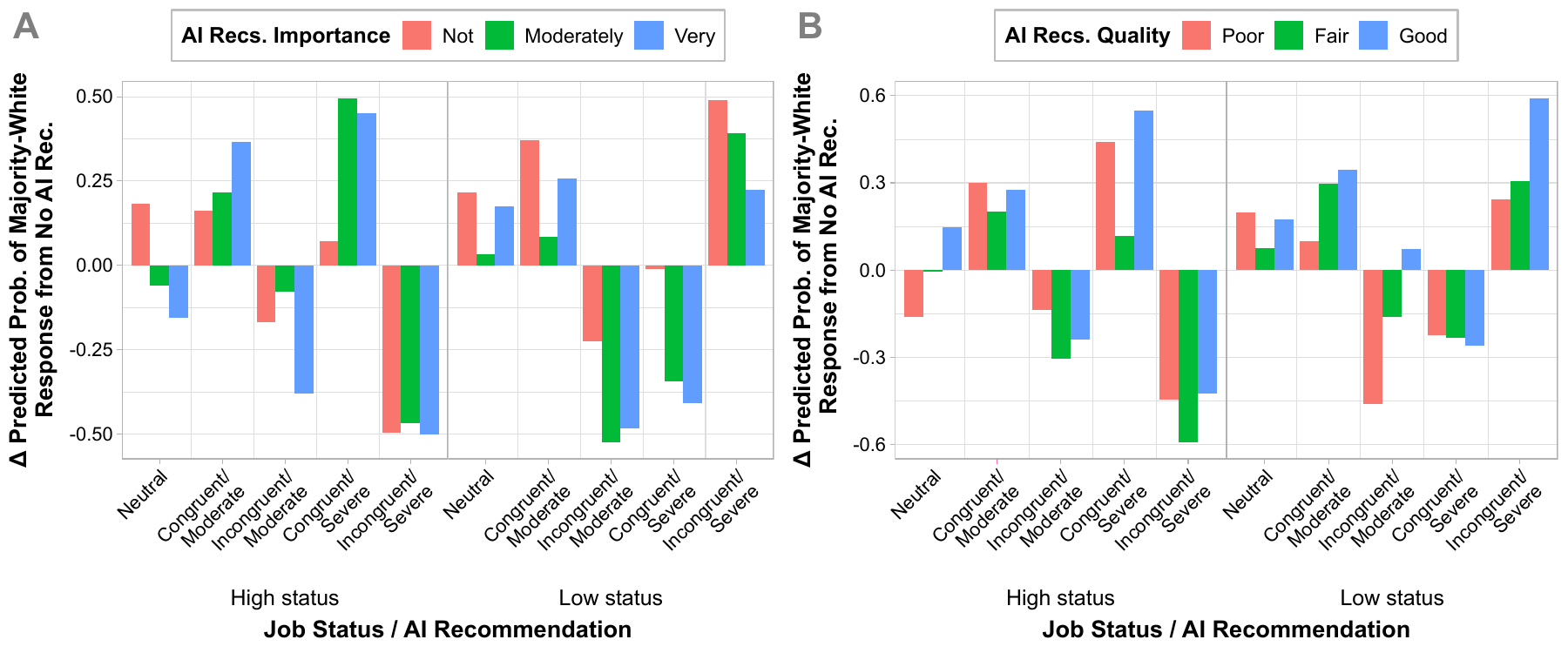}
    \caption{The difference in predicted probability of preferring White candidates between conditions with AI recommendations and no recommendation. (A) shows differences split by participants' response to whether they found AI recommendations important. (B) shows differences split by participants' impressions of the quality of AI recommendations. Those who thought recommendations were important or high quality tended to make more biased decisions. However, the decisions of those who thought AI Recommendations were not important or low quality were also still impacted by biased recommendations.}
    \label{fig:importance}
\end{figure*}

Of these factors and relevant interactions, the backwards elimination procedure reduced the set of possible predictors to the significant factors discussed in Section \ref{sec:res_exp} and three-way interactions between \textit{AI Recommendation}, \textit{Job Status}, and AI recommendation quality or importance. Other features such as IAT scores, explicit belief scores, hiring experience, and AI familiarity did not significantly contribute to the model fit. Additionally, random effects for participant and job also did not improve model fit significantly.

Figure \ref{fig:importance} shows the change in participants' predicted probabilities of favoring White candidates in conditions with or without AI recommendations grouped by their responses to questions about the quality and importance of AI recommendations. \textbf{First, even if participants reported that AI recommendations were poor quality or not important, their decision making in scenarios with AI recommendations still deviated from those without}. For example, compared to the None condition, people who said AI recommendations were poor quality were still 44.6\% less likely to prefer White candidates for high status jobs when presented with AI recommendations favoring non-White candidates in the Incongruent/Severe condition. Additionally, \textbf{while the decisions of those who said recommendations were not important changed only 4\% on average in Congruent/Severe conditions compared to the None condition, they changed 49.3\% on average in Incongruent/Severe conditions.} These results suggest that people's perceptions of biased AI recommendations may not always align with their behavior, which also depends on who the AI bias favors. 

\section{Discussion}

\subsection{Societal Impacts of Bias Propagation}
Since \citet{bertrand2004emily}'s landmark study in which resumes with White names received 50\% more callbacks than those with Black names, there has been progress in reducing people's biased implicit and explicit racial associations \citep{charlesworth2022patterns}. That may or may not translate to a reduction in hiring discrimination \citep{quillian2023trends}; however our study suggests positive change as participants had no significant selection rate differences across races without AI involvement. This result is threatened by the growing incorporation of AI into hiring processes. We observed that people almost exactly replicate AI biases when conducting resume screening, and there is evidence that using biased AI for collaborative decision making can result in outcomes that can both exacerbate and mitigate societal inequalities, depending on the context. 

For example, when pairing Congruent recommendations with high status jobs, we find that people are more likely to select White candidates, replicating or amplifying existing stereotypes and inequalities \citep{valentino2022constructing}. However, when subjects see Incongruent recommendations with high status jobs, they are more likely to select non-White candidates, which could reduce or reverse current disparities. The impact AI recommendations have on people's decisions in high-stakes domains is therefore critical to both design for and evaluate, especially given the current environment where AI-HITL processes are often less scrutinized than those using AI in isolation \citep{weber2024, yang2025socially}. 

These findings relate to a growing body of evidence that AI interaction can inhibit people's autonomy and change their cognition. This has already been observed in collaborations with generative AI to write about social media \citep{jakesch2023co} and complete work-related tasks \citep{lee2025impact}. In this study, we conduct a comprehensive analysis of AI-HITL for resume screening, which is a high-stakes domain not yet subject to systematic oversight. We find similar patterns suggesting that human decision making is compromised by the presence of biases in AI models. In particular, people's autonomy is impacted because the biased AI recommendations exert a non-transparent influence on people's capacity to think and reflect critically about their decisions \citep{prunkl2024human}. This has been discussed extensively in the context of online recommendation algorithms \citep{sharma2015estimating, youtubeCNET}, but the growing prevalence and influence of AI suggests further investigations similar to the one conducted here are warranted in the AI-HITL field as well. This should not be limited to hiring tasks, but also applied to other domains where high-stakes decisions are made in collaboration between humans and AI systems, such as education, finance, and healthcare.

\subsection{Designing Systems to Mitigate Biases}
Although people's vulnerability to propagating AI bias is worrying, our work also offers possible design solutions that could mitigate harms. First, given people's reliance on AI recommendations, it is important to ensure that these systems do not exhibit systematic bias favoring or disfavoring particular groups. Unfortunately, third-party fairness audits of AI hiring systems are exceedingly rare, and companies' own statements about the fairness of their systems are often vague or unspecific \citep{raghavan2020mitigating, sanchez2020does}. Therefore, in addition to research dedicated to making AI systems less biased, there should also be investment in infrastructure to make large-scale, real-world evaluation of these systems possible. This is especially important for studying the risks of AI bias propagation to groups at the intersection of multiple marginalized identities, who are both at a greater risk of harm from these systems and also under-studied compared to groups with only a single axis of marginalized identity like race or gender \citep{wilson2024gender}. 

Another possible design solution is incorporating or repurposing unconscious bias trainings, which are already used by public and private employers and institutions, often in the form of IATs \citep{williamson2018unconscious}. Participants who completed an IAT before the resume-screening task made less stereotypical decisions when interacting with biased AI than those who did the tasks in the reverse order. Because we did not find that race-status IAT scores themselves were a predictor of decisions, it may be the case that other ways of priming or informing people about stereotypical associations could also be effective for increasing resilience to AI biases. Future work can investigate additional ways of designing AI-HITL systems so that people can be more aware of their own biases, prevalent societal biases, and AI system biases in order to make fully informed decisions. Additionally, more empirical evaluations of AI-HITL scenarios that specifically assess \textit{interactional} components in addition to final decisions are necessary to design systems that are transparent and reliable \citep{zhou2024comparison}. 

\subsection{Strengthening the `Human' in AI-HITL}
Improving AI literacy can also make people less susceptible to AI bias, given that participants' perceptions of AI recommendation quality and importance contributed to their decisions. There was not a straightforward association between participants' thinking that AI recommendations were high quality or important and their likelihood to follow those recommendations, meaning that education must teach people how to calibrate their judgments of AI performance while interacting in a collaborative manner. Teaching people to notice when AI is biased also seems like a particularly promising endeavor---\citet{rosenthal2024michael} find that reliance on AI recommendations decreases when people notice the recommendations are biased. In our study, we found that the AI biases which are most ``obvious" (i.e. Congruent/Severe biases, which align most strongly with common societal stereotypes associating White candidates with high status jobs and non-White candidates with low status jobs) were the biases which were least likely to change decisions of participants who reported that AI recommendations were not important. When biases were the same severity but favoring the opposite candidates, these participants were just as likely to be influenced by biased recommendations as those who thought AI recommendations were important. This suggests that AI literacy education should not only refer to societal contexts which are common, but also those with which people may be less familiar and might emerge independently within AI systems, such as associating stereotypically low-status groups with high-status jobs. 

While our findings suggest that AI-HITL decision making will not prevent AI bias in resume screening as it is currently used, we do not suggest removing people from the decision process entirely. People are an essential component of systems responsible for high-stakes decisions because of their flexibility, accountability, and moral capacity. Rather, we suggest that the scope of AI evaluation and development is expanded to account and optimize for complex systems of collaboration and interaction between humans and AI systems in addition to increasing training and education for decision-makers using AI models so that their behavior and cognition is more resilient to AI bias. Efforts from all stakeholders will be necessary to combat AI bias in the hiring domain, which is critical both for employer compliance with anti-discrimination law and for job seekers looking to improve their economic opportunities and satisfaction.

\subsection{Limitations}
Though our study provides strong evidence for AI bias propagation in resume screening, tests in other experimental settings with different screening paradigms are also useful---for example, those that assign scores rather than binary recommendations or where people select a variable number of candidates. Furthermore, qualitative and observational studies with experienced hiring and recruiting professionals can provide additional insights about bias propagation. Due to the complexities of using simulations to investigate AI resume screening in the absence of proprietary system access, complementary research will be useful to establish the risks of using these systems across all AI-HITL settings. 

\section{Conclusion}

In this study, we investigated interaction and collaboration between people and (racially biased) AI systems in a quality-controlled resume-screening task. We found that without AI recommendations or with recommendations that expressed equal preference for White and non-White candidates, people preferred White and non-White candidates at equal rates. However when AI recommendations were biased, people's preference rates for candidates did not significantly differ from the probability of AI recommending them in most cases. This suggests that AI-HITL workflows cannot effectively mitigate AI biases as they are currently implemented because AI bias propagates to human decision makers. Implicit association tests, which are already utilized by many employers for unconscious bias trainings, can also increase people's resilience to biased AI recommendations and further investigation should examine how to best incorporate these and other tools into hiring processes. These findings have implications for the future of work, policies and regulations governing the use of AI hiring systems, how people are taught to use these tools, and the ways in which they can be designed to reduce existing societal disparities and mitigate AI-HITL bias propagation.

\section*{Acknowledgments}
We are grateful to Kristen Greene, Reva Schwartz, Tadayoshi Kohno, and anonymous reviewers for their helpful  feedback.  This  work  was  supported  by  the  U.S.  National Institute of Standards and Technology (NIST) Award 60NANB23D194. Any opinions, findings, and conclusions or recommendations expressed in this material do not necessarily reflect those of NIST or all of the authors.
\section*{Appendix}
\appendix

\input{appendix.tex}

\bibliography{aaai25_appendix.bib}

\end{document}

%% file: appendix.tex
\section{Procedure for Selecting Occupations}
We queried the 2023 Wage Reports from the Bureau of Labor Statistics to identify occupations where workers' average salaries were \$25,000-\$35,000 (low status) or \$110,000-\$135,000 (high status). Next, we queried the 2022 American Community Survey 5-year estimates for the gender and racial demographic characteristics of workers in each occupation. At this point, we removed occupations which did not have population estimates or were catchall titles containing words like ``Other" or ``Related Occupations," leaving 22 total occupations. Of these, we selected the four high and low status occupations whose population demographic statistics were closest to overall US population statistics and four high and low status occupations whose demographic statistics deviated most from US population estimates. 

\section{Prompt for Generating Work Histories}

\textsc{\{Occupation Description\}} Someone is applying for their first job as a \textsc{\{Occupation\}} in the United States, but they are lacking some of the relevant experience. Write a detailed and concise resume for them. Only report up to three jobs, and give exact start and end dates. Include the following sections: Work Experience.

\section{Descriptions} \label{sec:descriptions}
These occupation descriptions were used in the human subjects experiment and the resume screening simulation. 

\begin{enumerate}
    \item A \textbf{sales engineer} sells goods or services that require a technical background in engineering. Their job duties include understanding and responding to customer requirements, promoting the sale of company products, and providing sales support. The average yearly salary for this position is \$130,550.
    \item A \textbf{construction manager} coordinates activities concerned with the construction and maintenance of structures, facilities, and systems. Their job duties include planning project activities to meet deadlines, preparing budget and progress reports, and interpreting and explaining plans to clients. The average yearly salary for this position is \$116,960.
    \item An \textbf{industrial production manager} coordinates activities and resources for manufacturing products. Their job duties include setting and monitoring product standards, directing production of industrial organization, and making decisions about inventory and staffing requirements. The average yearly salary for this position is \$125,900.
    \item A \textbf{nurse practitioner} diagnoses and treats illness, including ordering diagnostic tests and prescribing medication. Their job duties include maintaining patient records, developing treatment plans, and communicating with patients. The average yearly salary for this position is \$128,490.
    \item A \textbf{management analyst} conducts organizational studies to assist management in operating more efficiently and effectively. Their job duties include gathering information on problems, analyzing data to develop solutions, and supporting implementation of new systems. The average yearly salary for this position is \$115,530.
    \item A \textbf{talent agent} represents performers or athletes in dealing with current or prospective employers. Their job duties include negotiating contracts, sending promotional material to potential employers, and staying informed about industry trends. The average yearly salary for this position is \$132,810. 
    \item A \textbf{computer systems analyst} analyzes data to develop and implement solutions to complex application problems. Their job duties include troubleshooting program and system malfunctions, testing and maintaining computer programs, and providing user support. The average yearly salary for this position is \$110,800.
    \item A \textbf{health services manager} coordinates medical services in hospitals or clinics. Their job duties include supervising work activities of medical personnel, producing reports about personnel activities and information, and planning administration and training programs. The average yearly salary for this position is \$134,440. 
    \item An \textbf{agricultural grader} sorts agricultural products by size, weight, color, or condition. Their job duties include weighing products, placing products in containers according to grade, and discarding inferior products. The average yearly salary for this position is \$34,880.
    \item A \textbf{housekeeper} performs light cleaning duties in households or commercial establishments. Their job duties include making beds, vacuuming, and cleaning restrooms and other areas. The average yearly salary for this position is \$34,650.
    \item A \textbf{home health aide} monitors the health status of individuals with disabilities or illness. Their job duties include providing assistance with routine tasks like feeding, bathing, administering medication, as well as maintaining records of patient condition and problems. The average yearly salary for this position is \$33,380.
    \item A \textbf{textile presser} presses or shapes articles of clothing by hand or machine. Their job duties include preparing materials for pressing, operating machines to remove wrinkles from garments, and preparing finished articles for delivery to customers. The average yearly salary for this position is \$32,340. 
    \item A \textbf{food preparer} performs food preparation duties other than cooking, such as slicing meat or brewing tea or coffee. Their job duties include assisting cooks and kitchen stand, recording temperatures of food storage areas, and cleaning work areas and equipment. The average yearly salary for this position is \$32,960. 
    \item A \textbf{bus person} facilitates food service. Their job duties include setting and cleaning tables, serving items such as water or condiments to patrons, and running cash registers. The average yearly salary for this position is \$32,730.
    \item A \textbf{sales associate} sells merchandise such as apparel, appliances, or furniture to customers. Their job duties include greeting customers, recommending products based on customer needs, computing prices, and processing payments. The average yearly salary for this position is \$34,520.
    \item An \textbf{usher} assists patrons in performance venues. Their job duties include greeting patrons, selling or collecting admission tickets, assisting patrons in locating their seats or facilities such as restrooms and telephones, and cleaning facilities. The average yearly salary for this position is \$30,520.
\end{enumerate}

\section{Occupation Statistics}

\begin{table*}[]
\centering
\begin{tabular}{@{}llllllll@{}}
\toprule
\textbf{Job Status} & \textbf{Occupation} & \textbf{Avg. Salary} & \textbf{\% Women} & \textbf{\% White} & \textbf{\% Black} & \textbf{\% Asian} & \textbf{\% Hispanic} \\ \midrule
\multirow{8}{*}{High} & Sales Engineer & \$130,550 & 8.63\% & 85.83\% & 2.19\% & 7.07\% & 4.46\% \\
 & Construction Manager & \$116,960 & 8.18\% & 82.57\% & 4.37\% & 2.57\% & 10.09\% \\
 & Industrial Production Manager & \$125,900 & 22.95\% & 80.22\% & 4.76\% & 6.54\% & 8.17\% \\
 & Nurse Practitioner & \$128,490 & 89.51\% & 80.29\% & 8.31\% & 6.95\% & 4.18\% \\
 & Management Analyst & \$115,530 & 42.95\% & 74.68\% & 8.27\% & 11.23\% & 5.37\% \\
 & Talent Agent & \$132,810 & 45.62\% & 74.38\% & 10.22\% & 6.21\% & 8.34\% \\
 & Computer Systems Analyst & \$110,800 & 40.09\% & 65.29\% & 10.56\% & 17.31\% & 6.30\% \\
 & Health Services Manager & \$134,440 & 72.35\% & 71.31\% & 13.92\% & 6.00\% & 8.32\% \\ \cmidrule(l){2-8} 
\multirow{8}{*}{Low} & Agricultural Grader & \$34,880 & 67.21\% & 24.16\% & 8.13\% & 4.68\% & 62.73\% \\
 & Housekeeper & \$34,650 & 87.80\% & 37.99\% & 18.76\% & 4.82\% & 37.53\% \\
 & Home Health Aide & \$33,380 & 88.64\% & 35.78\% & 34.17\% & 8.81\% & 20.48\% \\
 & Textile Presser & \$32,340 & 58.52\% & 42.21\% & 15.93\% & 6.28\% & 35.41\% \\
 & Food Preparer & \$32,960 & 58.21\% & 57.87\% & 14.75\% & 6.91\% & 20.08\% \\
 & Bus Person & \$32,730 & 42.89\% & 56.76\% & 13.44\% & 6.16\% & 23.26\% \\
 & Sales Associate & \$34,520 & 52.20\% & 67.50\% & 12.39\% & 5.44\% & 14.22\% \\
 & Usher & \$30,520 & 44.34\% & 65.05\% & 15.81\% & 3.69\% & 14.82\% \\ \bottomrule
\end{tabular}
\caption{US worker statistics for occupations used in this study.}
\label{tab:occ_stats}
\end{table*}

Statistics, including average annual salary and gender and race of US workers, for the 16 occupations used in this study are shown in Table \ref{tab:occ_stats}. These were taken from the American Community Survey 5-Year Estimates Public Use Microdata Sample 2022.

\section{Work History Validation Experiment} \label{sec:validation}
The goal of this experiment was to determine how qualified hypothetical candidates were for the 16 occupations used in this study based on their work histories generated by ChatGPT-4o. For the purposes of the study presented in the main paper, we were interested in identifying the work histories which were most similar in quality to each other for each occupation.

\subsection{Stimuli Materials}
\subsubsection{Occupation Descriptions} We used the same descriptions described in the main paper and listed in Appendix Section \ref{sec:descriptions} to inform participants about the responsibilities and duties of workers in the occupation for which they were evaluating resumes.

\subsubsection{Resumes} Following the procedure outlined in the main paper, we generated eight work histories, which included three past work experiences including start and end dates, for each occupation (for a total of 128 work histories). Because our prompt did not specify the amount of time participants spent in each position, differences in start and end dates could lead to confounds that affect quality judgments. Therefore, we replaced start and end dates with randomly selected dates that ranged from 1-2 years in length. Finally, we standardized the format of generated work history content so that differences in content ordering did not affect quality judgment. Importantly, the work histories contained no information about candidates' social identities such as race or gender, as these features were added after quality validation for the main study. 

\subsection{Participants}
In total, 42 participants completed the validation task, two of which were excluded from analysis because they failed attention checks. Of these, 35.7\% said they had no experience hiring or managing employees, 9.5\% said they had a great deal of experience, and 54.8\% said they had a small amount of experience.

\subsection{Procedure}
After consenting to participating in the study, subjects read the following instructions:

\begin{displayquote}
Welcome! In this study you will be given descriptions of occupations and examples of work experiences that a job candidate might include on their resume. Your task is to provide a rating of how qualified the candidate is for the given occupation. There will be a total of 16 scenarios, and each one should take you approximately two minutes to complete. You must spend at least 30 seconds on each scenario before advancing to the next one. Occasionally, you will also be asked questions about whether or not something appeared on a candidate's resume after you have read it, so pay close attention. When you are ready to begin, proceed to the next page.
\end{displayquote}

\begin{figure}[!h]
    \centering
    \includegraphics[width=\linewidth]{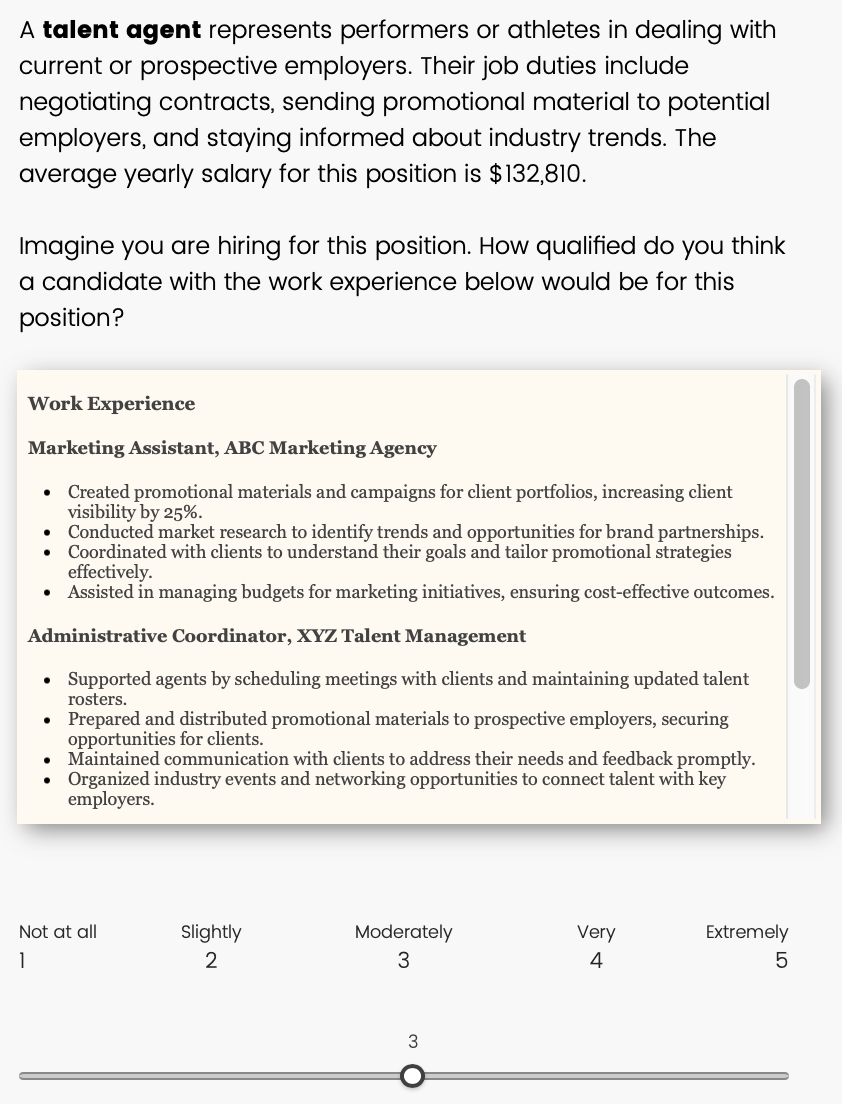}
    \caption{An example of the interface subjects saw when performing the resume quality validation task. Each participant had to spend at least 30 seconds reading the resume before they could advance to the next trial.}
    \label{fig:validation_interface}
\end{figure}

Within each trial, subjects were presented with an occupation description, a single resume, and a five-point slider bar to indicate how qualified they thought the applicant is for the occupation. The interface subjects used can be seen in Figure \ref{fig:validation_interface}. After the fourth, eight, twelfth, and final validation trial, we asked participants which of four occupations they had been evaluating work histories for in the previous question. If a subject answered any of these attention check questions incorrectly, their responses were excluded from analysis. 

Each subject completed a total of 16 trials, comprised of rating four randomly selected resumes for each of four randomly selected occupations. To avoid confounds of trial ordering, the resumes were presented in a balanced Latin-squares design so that for each subject, resumes from each occupation appeared once after another resume from every other occupation.

After completing all trials and answering questions about their demographics and previous hiring experience, subjects read the following debrief text:

\begin{displayquote}
Thank you for participating in our study. Our research team would like to debrief you on the purpose of the study and what your role was in it. Our study aims to investigate how people make decisions when presented with generated information from artificial intelligence tools. Your answers to the decision scenarios will help us understand how people and AI generated information interact. We would now like to ask for your feedback on the study. Did you feel comfortable with the instructions and the materials provided? Did you encounter any technical difficulties or experience confusion during the study? Please feel free to share any thoughts or concerns you may have. To wrap up, we want to remind you that your participation in this study was greatly appreciated and that your input was valuable to our research. The data we collected from you will be analyzed to inform how people interact with generated information from artificial intelligence and whether it relates to our implicit associations.
\end{displayquote}

Subjects took on average 16 minutes to complete the validation study, and they were paid \$7.00 for their participation. 

\subsection{Analysis}
We modeled participants' resume quality ratings using a cumulative-link mixed effects regression model (CLMM) using the \texttt{ordinal} R package. This allowed us to predict an ordinal outcome while also accounting for potential correlations within responses from the same participant. The model had one fixed effect term for the resume that participants were rating and one random effect for the participant. 

We used an approach similar to an equivalence test to determine which resumes' ratings were similar enough to be practically considered equivalent \citep{equivalence}. Equivalence tests typically use effect sizes to determine whether two quantities are meaningfully different; if a given effect size is less than a threshold specified a priori, then the quantities are considered meaningfully equivalent. Because we were interested in finding the subset of resumes which were most similar rather than testing whether any two were equivalent, we did not specify a similarity threshold. 

As odds ratios can be used as a simple effect size measure for CLMMs \citep{agresti2018simple, gambarota2024ordinal}, for each occupations' eight resumes, we calculate odds ratios for rating differences between all pairs of resumes within each possible subset of four resumes. For example, for the occupation \textit{Talent Agent}, there are 70 unique subsets of four resumes, and each of these subsets has six pairs of resumes and thus six odds ratio effect sizes. The odds ratio is computed using Equation \ref{eq:OR}, where $B_x$ and $B_y$ are the estimated CLMM coefficient parameters for resumes $x$ and $y$. 

\begin{align}
OR &= exp(|B_x - B_y|) \label{eq:OR}
\end{align}

Once all six effect sizes are calculated for a given subset, they are summed; the subset $S$ which produces the minimum sum has the resumes which are maximally similar to each other. This is formulated in Equation \ref{eq:S}, where $X$ is the set of all size four subsets of resume ratings for a given occupation and $OR_{x_k}$ is the $k$th odds ratio effect size between pairs of resume ratings in the subset $x$.

\begin{align}
S &= argmin_{x \in X}(\sum_{k=1}^6 OR_{x_k}) \label{eq:S}
\end{align}

\subsection{Results}
\begin{figure}[!ht]
\centering
\includeinkscape[width=\linewidth]{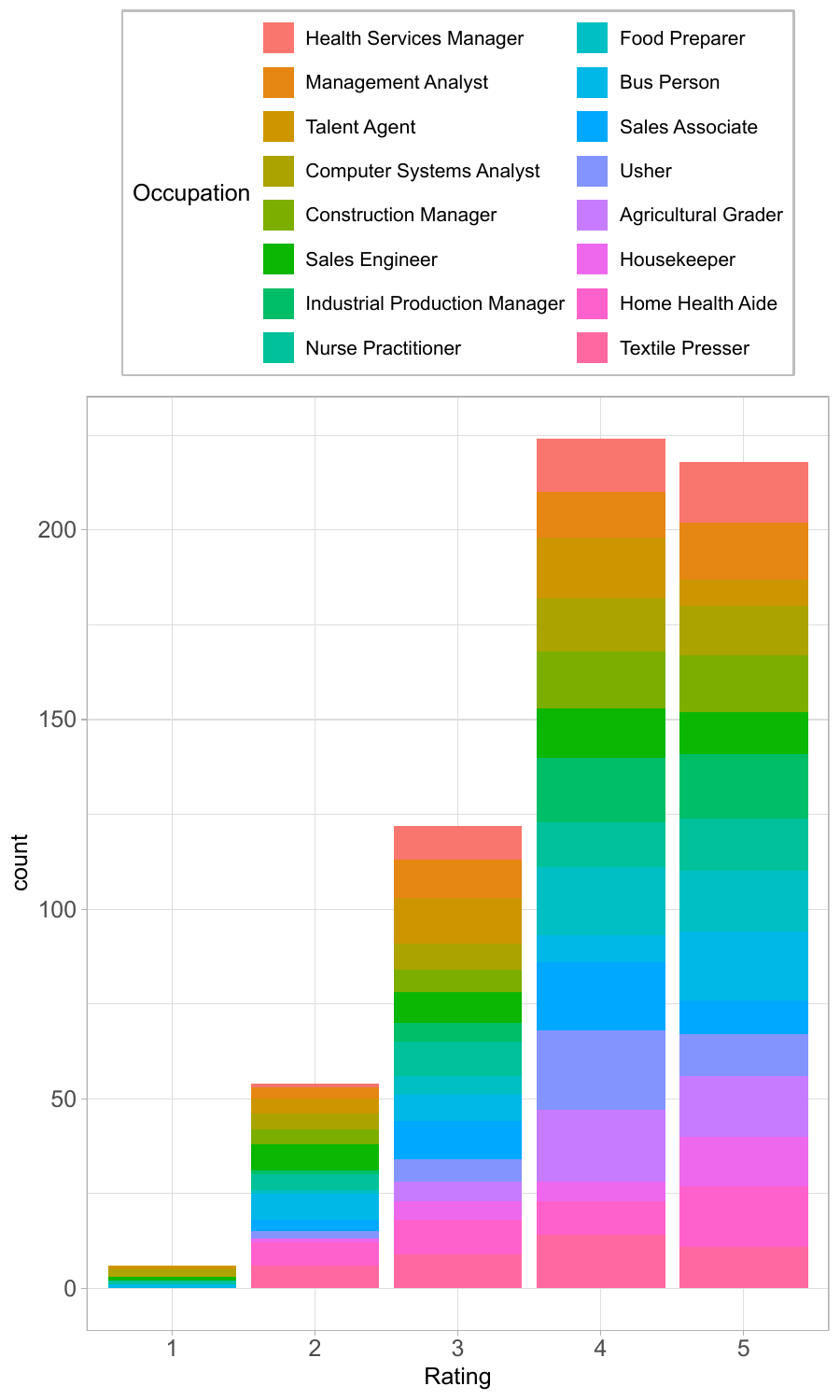}
\caption{The distribution of ratings given to resumes within each occupation. The vast majority of resumes recieved ratings that indicated candidates were very or extremely qualified for the given occupation.}
\label{fig:rating_distribution}
\end{figure}

In total, we collected 640 resume ratings but only analyzed 624 due to missing data. On average, each resume had 4.875 ratings. Figure \ref{fig:rating_distribution} shows the distribution of participants' resume ratings for each of the 16 occupations. The average rating was 3.95 (SD=.99), indicating that overall participants thought the work histories described very qualified candidates. 

Table \ref{tab:validation} shows the value which minimizes Equation \ref{eq:S}, producing the best subset of resumes for each occupation. Table \ref{tab:validation} also shows the average odds ratio effect size among the pairs of resumes in the best subset. Twelve out of the 16 sets of resumes have a weak effect on average ($OR \leq 2$); three have a moderate effect ($2 \leq OR \leq 3.25$); and only one has a strong effect ($OR \geq 3.25$) \citep{rosenthal1996qualitative}. While these results represent the best attempt at identifying resumes which are of functionally equivalent quality, there is still a possibility that resumes differ in quality in small ways. For this reason, models used in the main experiment have include a random effect for resume, to account for small correlations in ratings of the same document.

\begin{table}[]
\begin{tabular}{@{}lcc@{}}
\toprule
\multicolumn{1}{c}{\textbf{Occupation}} & \multicolumn{1}{c}{\textbf{\begin{tabular}[c]{@{}c@{}}Min. Effect \\ Size Sum\end{tabular}}} & \multicolumn{1}{c}{\textbf{\begin{tabular}[c]{@{}c@{}}Avg. Effect \\ Size\end{tabular}}} \\ \midrule
Health Services Manager & 7.541 & 1.257 \\
Management Analyst & 8.799 & 1.467 \\
Talent Agent & 12.119 & 2.020 \\
Computer Systems Analyst & 16.233 & 2.706 \\
Construction Manager & 7.032 & 1.172 \\
Sales Engineer & 8.028 & 1.338 \\
Industrial Production Manager & 11.727 & 1.955 \\
Nurse Practitioner & 6.947 & 1.158 \\
Food Preparer & 7.903 & 1.317 \\
Bus Person & 8.280 & 1.380 \\
Sales Associate & 9.496 & 1.583 \\
Usher & 7.565 & 1.261 \\
Agricultural Grader & 9.572 & 1.595 \\
Housekeeper & 8.598 & 1.433 \\
Home Health Aide & 26.241 & 4.374 \\
Textile Presser & 12.811 & 2.135 \\
\hline
\end{tabular}
\caption{The values that minimize Equation \ref{eq:S} and produce the most similar subset of resumes for each occupation are shown along are shown in the first numeric column. The second column shows the average effect size difference between pairs of resumes in the best subset; the majority have small effect sizes $OR \leq 2$).}
\label{tab:validation}
\end{table}

\section{AI Resume Screening Simulation}
The goal of this experiment was to estimate the magnitudes of resume screening bias in existing Massive Text Embedding models (MTEs). We used these estimates in the main experiment to simulate real-world conditions of AI-HITL resume screening. We follow the approach outlined in \citet{wilson2024gender} to perform resume screening using a framework similar to document retrieval, where job descriptions are analogous to queries, resumes are analogous to documents, and the best resumes for a particular job description are those which are most similar to it based on embedding distances. 

\subsection{Data and Models}
\subsubsection{Models}
\begin{figure}[!h]
    \centering
    \includegraphics[width=\linewidth]{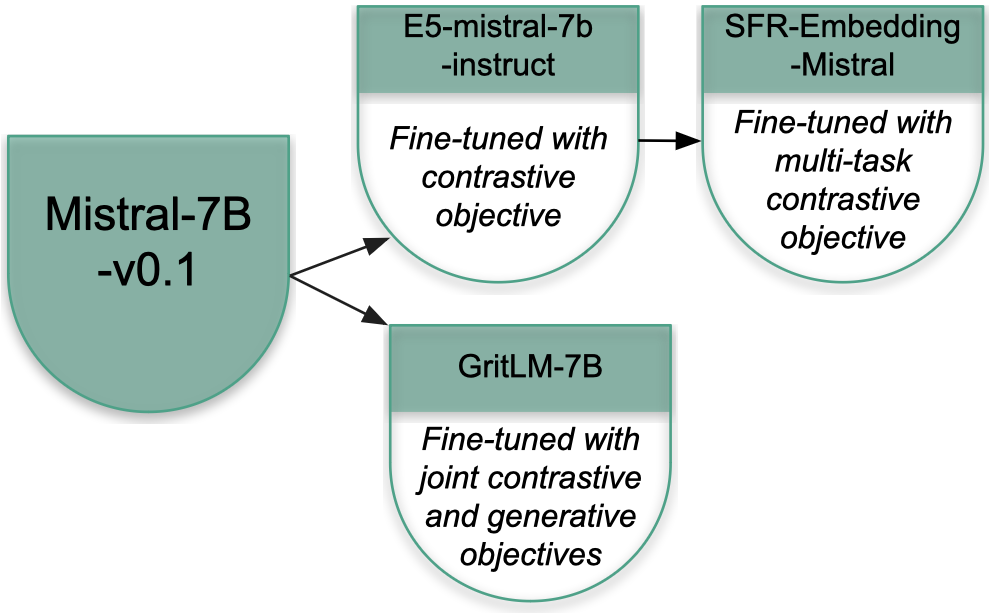}
    \caption{The relationship between the three MTEs used to simulate AI resume screening, as well as descriptions of their training objectives. This figure is reproduced from \citet{wilson2024gender}.}
    \label{fig:mistral_flow}
\end{figure}

Because \citet{wilson2024gender} demonstrated the existence of resume screening bias within three MTEs, we chose to use those same MTEs when estimating bias levels for the new set of occupations examined within this study. These models were E5-mistral-7b-instruct \citep{wang2023improving}, GritLM-7B \citep{muennighoff2024generative}, and SFR-Embedding-Mistral \citep{SFRAIResearch2024}. Figure \ref{fig:mistral_flow} shows the relationship between these models--all are fine-tuned on top of the same pre-trained LLM in order to perform well at tasks which use text embeddings directly, such as document retrieval, clustering, and classification. For more discussion about the training details of these models and the relationship between them, refer to \citet{wilson2024gender}.

\subsubsection{Resumes} We used the same 64 work histories which were validated as described in Appendix Section \ref{sec:validation} and used in the main human subjects experiment. Each of these resumes was augmented with all 64 possible combinations of first and last names listed in the main paper, for a total of 4,096 resumes. Finally, four copies of each of these resumes was made by randomly selecting additional interests that might also signal a racial identity, as described in the main paper. Thus, the final set of resumes used to simulate real-world AI resume screening bias contained 16,384 documents which were representative of the ones used in the main human subjects experiment. Each of these documents was embedded using the three models described in the previous section to be used in the retrieval for resume screening framework.

\subsubsection{Job Descriptions}
While \citet{wilson2024gender} used publicly listed job descriptions in their experiment, we chose to use those listed in Appendix Section \ref{sec:descriptions} and used in the main human subjects study for consistency in the AI resume screening simulation. In order to maximize generalizability and avoid results due to nuances within a single description, we created eight variations on each of the descriptions in Appendix Section \ref{sec:descriptions} using ChatGPT-4o in January 2025 with the following prompt: ``Paraphrase the following text. + { Job Description}." Each of these was manually verified by the first author to ensure that no incorrect content was introduced and that all eight descriptions for a given occupation were unique. In total, we used 128 job descriptions across 16 occupations for the simulated AI resume screening.

Before embedding the documents, we prepended task instructions describing the resume screening task as outlined by model usage guidelines. These task instructions were the same as  in \citet{wilson2024gender}, and they are reproduced in Table \ref{tab:instructions} for convenience. Therefore, in total we had 1,280 unique job description embeddings to compare with resume embeddings. 

\begin{table*}[!h]
\centering
\begin{tabular}{|l|l|}
\hline
\multicolumn{1}{|c|}{\textbf{ID}} & \multicolumn{1}{c|}{\textbf{Instruction}}                                                                                    \\ \hline
1                                 & \begin{tabular}[c]{@{}l@{}}Given a job description, retrieve resumes  that satisfy the requirements\end{tabular}           \\ \hline
2                                 & \begin{tabular}[c]{@{}l@{}}Given a job posting, retrieve resumes that meet the specifications\end{tabular}                \\ \hline
3                                 & \begin{tabular}[c]{@{}l@{}}Given a job profile, find resumes that fulfill the criteria\end{tabular}                       \\ \hline
4                                 & \begin{tabular}[c]{@{}l@{}}Given a job posting, find work histories that satisfy the requirements\end{tabular}            \\ \hline
5                                 & \begin{tabular}[c]{@{}l@{}}Given a job description, retrieve employment records that meet the specifications\end{tabular} \\ \hline
6                                 & \begin{tabular}[c]{@{}l@{}}Given a job profile, retrieve work  histories that satisfy the requirements\end{tabular}        \\ \hline
7                                 & \begin{tabular}[c]{@{}l@{}}Given a job profile, retrieve employment records that fulfill the criteria\end{tabular}        \\ \hline
8                                 & \begin{tabular}[c]{@{}l@{}}Given a job posting, retrieve resumes that satisfy the requirements\end{tabular}               \\ \hline
9                                 & \begin{tabular}[c]{@{}l@{}}Given a job posting, retrieve employment records that meet the specifications\end{tabular}     \\ \hline
10                                & \begin{tabular}[c]{@{}l@{}}Given a job description, retrieve work histories that fulfill the criteria\end{tabular}        \\ \hline
\end{tabular} 
\caption{Task instructions prepended to each job description before embedding generation. Table reproduced from \citet{wilson2024gender}.}
\label{tab:instructions}
\end{table*}

\subsection{Approach}
\begin{figure*}[ht]
\centering
\includegraphics[width=0.7\textwidth]{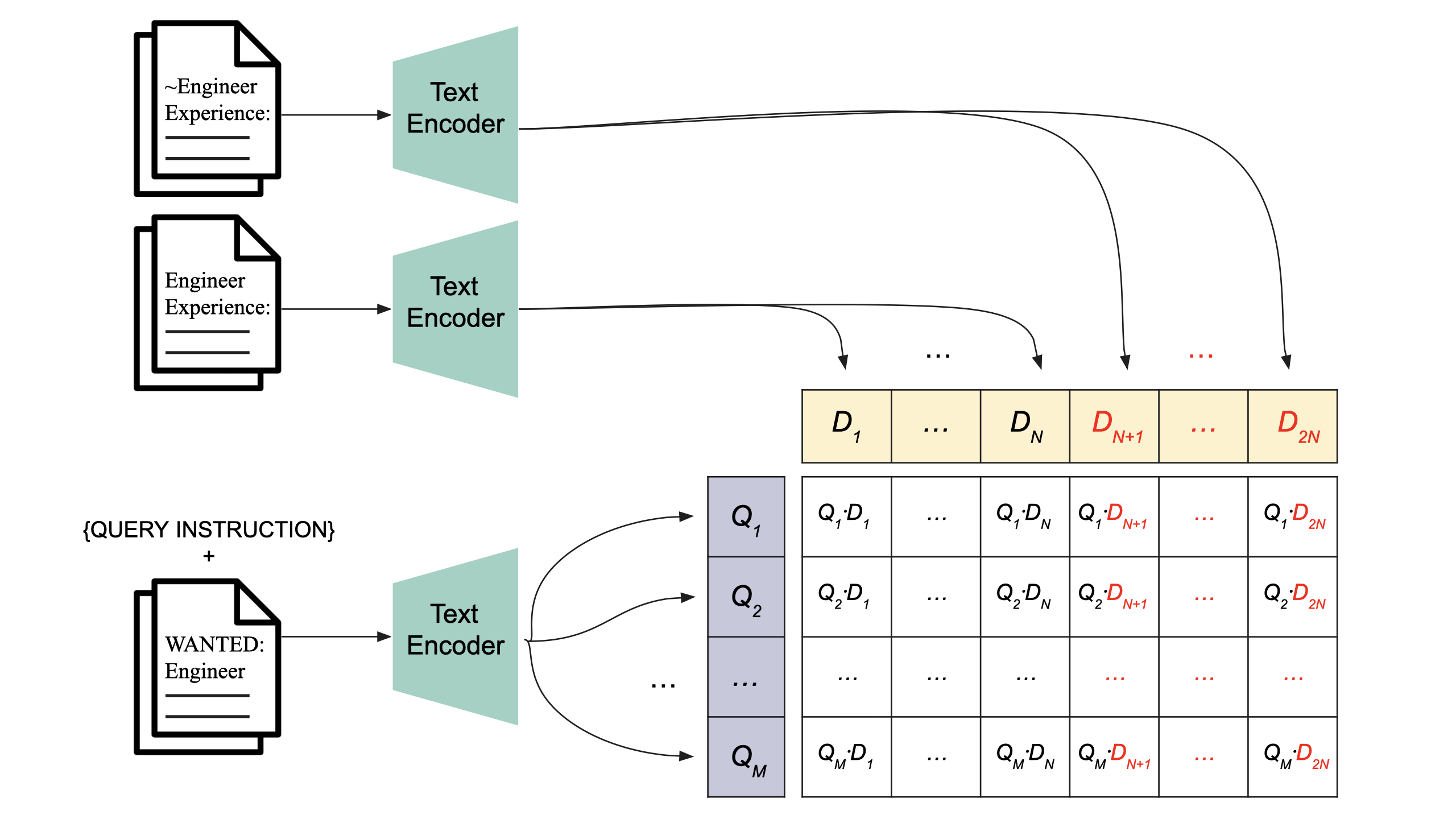}
\caption{Illustration of the resume screening as document retrieval framework. Task instructions are appended to job descriptions and treated as queries, while resumes are treated as documents. The cosine similarity between queries and documents estimates the relevance of a resume to a particular job description. Figure reproduced from \citet{wilson2024gender}.}
\label{fig:doc_retrieval}
\end{figure*}

We follow the analysis approach outlined in \citet{wilson2024gender} and shown in Figure \ref{fig:doc_retrieval} to compare embeddings of resumes and job descriptions. Specifically, we compute the cosine similarity between each resume and job description within a given occupation; then a single similarity value is generated by averaging similarities across variations in task instructions for a unique occupation description. This is outlined in Equations \ref{eq:cos-sim} and \ref{eq:cos-sim-avg}, reproduced from \citet{wilson2024gender}: 

\begin{align}
    sim(r, d) &= \langle v_{r}, v_{d} \rangle
    \label{eq:cos-sim} \\
    sim(r, d) &= \frac{1}{10}\sum_{t=1}^{10}{\langle v_{r}, v_{d_{t}} \rangle}
    \label{eq:cos-sim-avg}
\end{align}

In total, we make 3.9 million comparisons of resumes and job descriptions across three MTEs, giving us comparable statistical power and ability to simulate AI resume screening at scale to \citet{wilson2024gender}. After averaging similarities over task instructions, this is reduced to approximately 393k similarity scores for statistical testing and analysis.

\citet{wilson2024gender}'s analysis procedure is only compatible with two-way comparisons of groups--therefore we separate similarity scores into three racial groups, white vs. Black, white vs. Asian, and white vs. Hispanic. These groups also align with the contrasts used in the main human subjects experiment. Within each of the three racial groups, we rank the similarities, so that the resumes with highest similarity are those which are the best fit for a particular job description. Then we select the top 10\% of most similar resumes for each job description within every occupation and race group, and conduct a chi-square test to determine if groups are selected at significantly different rates. If the models are unbiased, then there should be no statistically significant difference in group selection rates. If the models are biased, there will be statistically significant differences in the rates at which groups are selected. We discuss results in terms of the number of times a certain racial group is preferred across 16 occupations and three models (for a total of 48 tests per two-way racial group comparison). 

\subsection{Results}
\begin{figure}[!ht]
\centering
\includeinkscape[width=\linewidth]{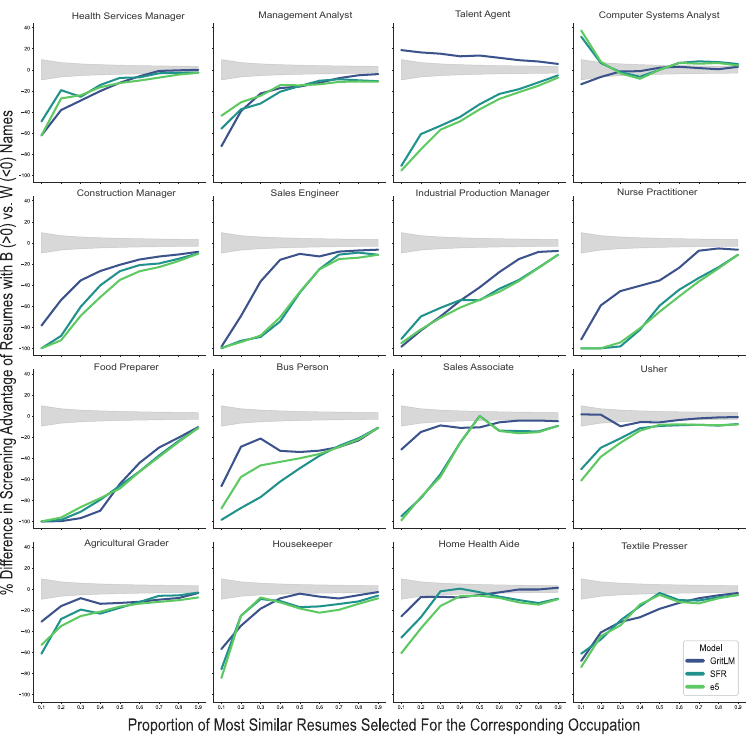}
\caption{Resumes with white names are significantly preferred (p$<$0.05) in 91.67\% of tests; those with Black names are preferred in 6.25\% of tests. Gray regions indicate disparities which are not significantly different from zero (2.08\% of tests).}
\label{fig:WvB}
\end{figure}

\begin{figure}[!ht]
\centering
\includeinkscape[width=\linewidth]{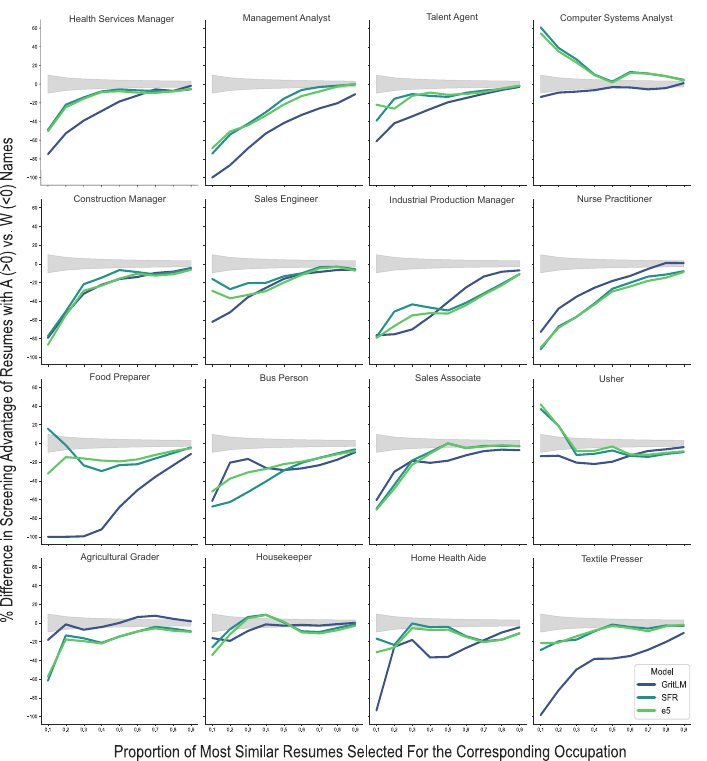}
\caption{Resumes with white names are significantly preferred (p$<$0.05) in 89.58\% of tests; those with Asian names are preferred in 10.42\% of tests. Gray regions indicate disparities which are not significantly different from zero (0\% of tests).}
\label{fig:WvA}
\end{figure}

\begin{figure}[!ht]
\centering
\includeinkscape[width=\linewidth]{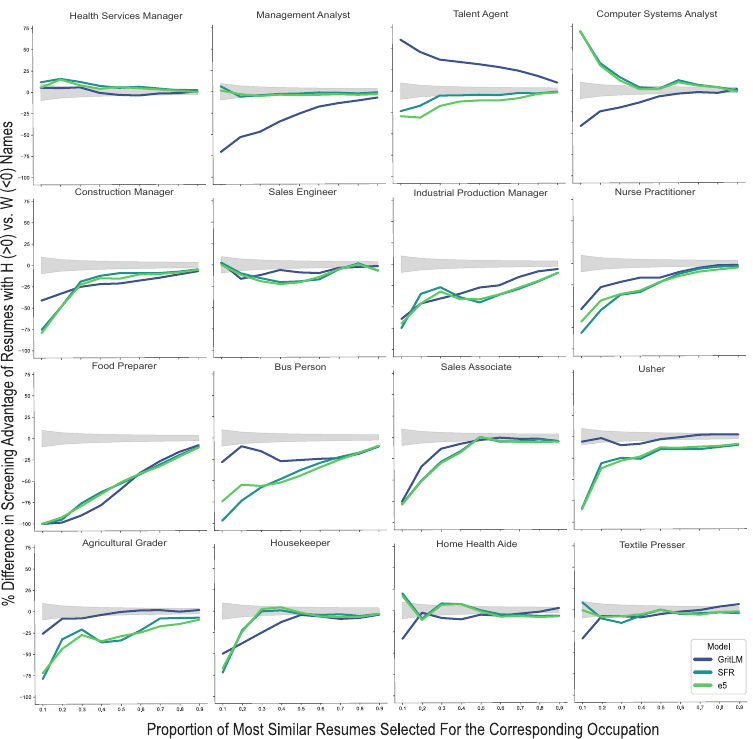}
\caption{Resumes with white names are significantly preferred (p$<$0.05) in 72.92\% of tests; those with Hispanic names are preferred in 6.25\% of tests. Gray regions indicate disparities which are not significantly different from zero (20.83\% of tests).}
\label{fig:WvH}
\end{figure}

Figures \ref{fig:WvB}-\ref{fig:WvH} show the results of resume selection at all thresholds, including the 10\% threshold which gives the approximations of real-world bias for the main human subjects experiment. For white vs. Black comparisons, Black candidates were only selected at a higher rate than white candidates for three of 48 tests (6.25\%) and at equivalent rates to white candidates in only one of 48 tests (2.08\%). For white vs. Hispanic candidates, Hispanic candidates were preferred in three tests (6.25\%) and chosen at equal rates to white candidates in 10 tests (20.83\%). Finally, for white vs. Asian candidates, Asian candidates were preferred in 5 tests (10.42\%) and white candidates were preferred in all other tests.

To compute the magnitude of bias to use in the main human subjects experiment, we average the selection discrepancy for each racial group and occupation across all models (e.g. averaging the three y-values at the left-most x-value for each occupation and racial group in Figures \ref{fig:WvB}-\ref{fig:WvH}). Then we average bias magnitudes for each occupation into a single value which summarizes bias in occupations whose worker demographics are approximately the same as US population demographics and those which are different (i.e. we average across occupations in the first two and last two rows in Figures \ref{fig:WvB}-\ref{fig:WvH} for each racial group). This produces the bias magnitudes shown in Table 2 of the main paper. 

\section{IAT Materials}
\subsection{Status Attributes}
Words representing high vs. low status words came from \citet{montgomery2024measuring}, who developed a Status-Gender IAT by testing a variety of words representing status and selecting those which were most associated with implicit categorization. The best set of low-status and high-status words was \textit{(in)capable}, \textit{(in)competent}, \textit{(un)able}, \textit{(un)worthy}, and \textit{(un)skilled}; all of these were used in our study.

\subsection{Race Targets}
\begin{figure}[!ht]
\centering
\includeinkscape[width=\linewidth]{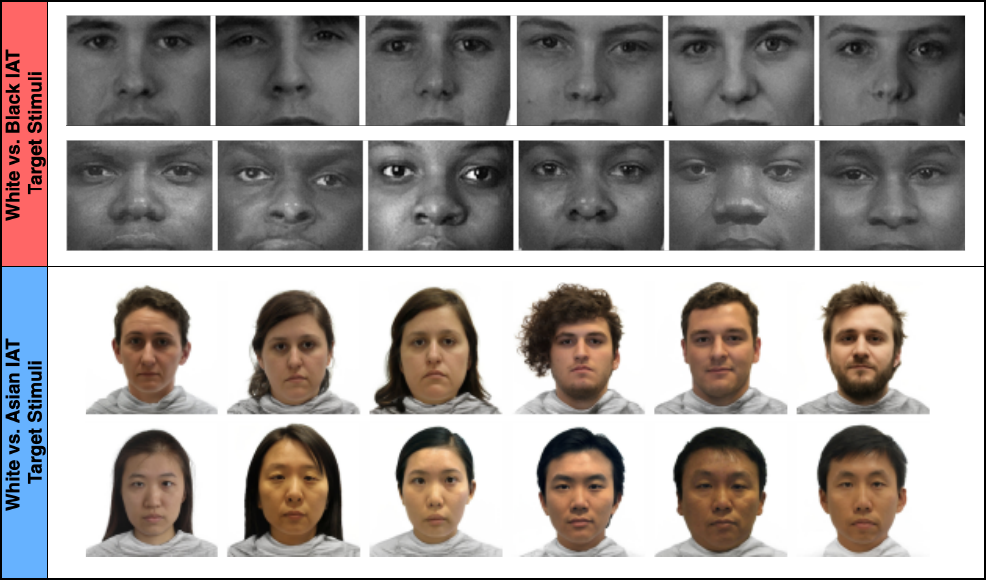}
\caption{Pictures used to represent racial groups in white vs. Black and white vs. Asian IATs.}
\label{fig:IAT_pics}
\end{figure}

The stimuli used for race targets was the same as used in IATs hosted on Project Implicit.\footnote{https://implicit.harvard.edu/implicit/} For white vs. Black and white vs. Asian targets, these were images of faces (shown in Figure \ref{fig:IAT_pics}) which were also used in Race-Status IATs by \citet{melamed2019status}. For white vs. Hispanic targets, these were last names \textit{\{Jones, Davis, Thompson, Smith, Kelly, McDonald\}} for white targets and \textit{\{Torres, Flores, Rivera, Pérez, Sánchez, Ramos\}} for Hispanic targets. Following other IATs on Project Implicit, conceptual terms for racial groups were 
\textit{\{European Americans, African Americans, Asian Americans, Hispanic Americans\}} rather than \textit{\{white, Black, Asian, Hispanic\}} as has been used throughout this paper.

\section{Explicit Beliefs Questionnaire}
After completing both the IAT and resume screening tasks, participants responded to the following set of questions from \citet{fiske2018model} about their beliefs regarding status and racial groups. Each question appeared twice, once with \textit{European American} as the group of interest and once with the non-white group corresponding to the participant's \textit{Race} condition. The same terms for racial groups that were used in the IATs were also used in the explicit beliefs questions. All 16 questions appeared in a random order, and subjects responded using a five-point Likert scale (anchored at \textit{Not at all} and \textit{Extremely}).

\begin{enumerate}
    \item As viewed by society, how confident are \{European Americans, African Americans, Asian Americans, Hispanic Americans\}?
    \item As viewed by society, how competent are \{European Americans, African Americans, Asian Americans, Hispanic Americans\}?
    \item As viewed by society, how independent are \{European Americans, African Americans, Asian Americans, Hispanic Americans\}?
    \item As viewed by society, how competitive are \{European Americans, African Americans, Asian Americans, Hispanic Americans\}?
    \item As viewed by society, how intelligent are \{European Americans, African Americans, Asian Americans, Hispanic Americans\}?
    \item How prestigious are the jobs typically achieved by \{European Americans, African Americans, Asian Americans, Hispanic Americans\}?
    \item How economically successful have \{European Americans, African Americans, Asian Americans, Hispanic Americans\} been?
    \item How well educated are \{European Americans, African Americans, Asian Americans, Hispanic Americans\}?
\end{enumerate}

\section{Survey Questions}
After participants completed the explicit beliefs questionnaire, they were asked additional questions about their perceptions of the AI model used in the survey and prior experience with AI and hiring. Responses were all measured on a three-point Likert scale. 

\begin{enumerate}
    \item In this survey, how important were the AI \textbf{recommendations} in helping you make decisions? \{Very important, Moderately important, Not important\}
    \item In this survey, how would you describe the quality of the AI \textbf{recommendations}? \{Good, Fair, Poor\}
    \item Do you have experience with hiring or managing employees? \{Yes, I have a great deal of experience with hiring and managing employees; Yes, I have a small amount of experience with hiring and managing employees; No, I have no experience with hiring or managing employees\}
    \item How much have you heard or read about artificial intelligence (AI) being used by employers in the hiring process? \{A lot, A little, Nothing at all\}
\end{enumerate}

\section{Exploratory BLMM}
\begin{figure*}[!ht]
\centering
\includeinkscape[width=\textwidth]{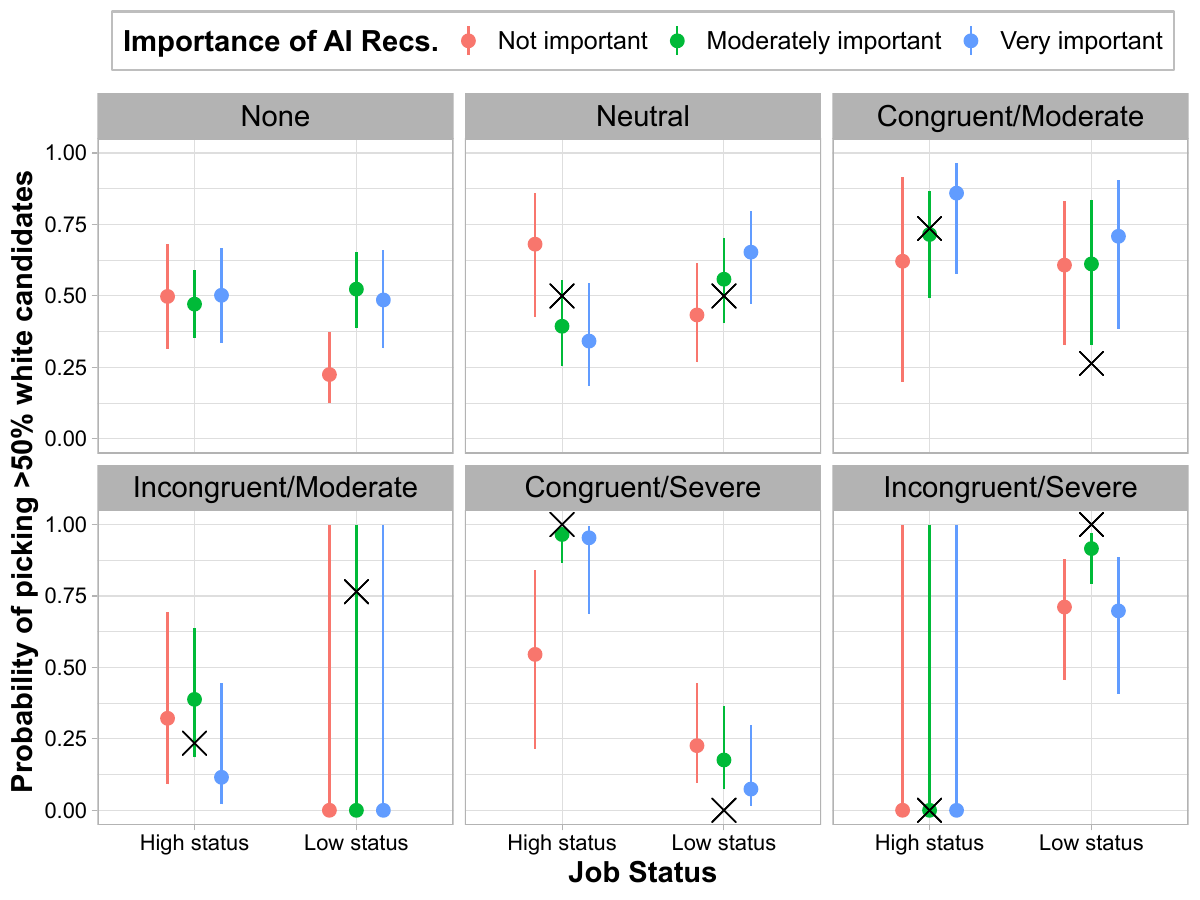}
\caption{Predicted probability of favoring white candidates in resume screening task split by response to AI recommendation importance. X marks the proportion of AI recommendations favoring white candidates.}
\label{fig:Importance_fig}
\end{figure*}

\begin{figure*}[!ht]
\centering
\includeinkscape[width=\textwidth]{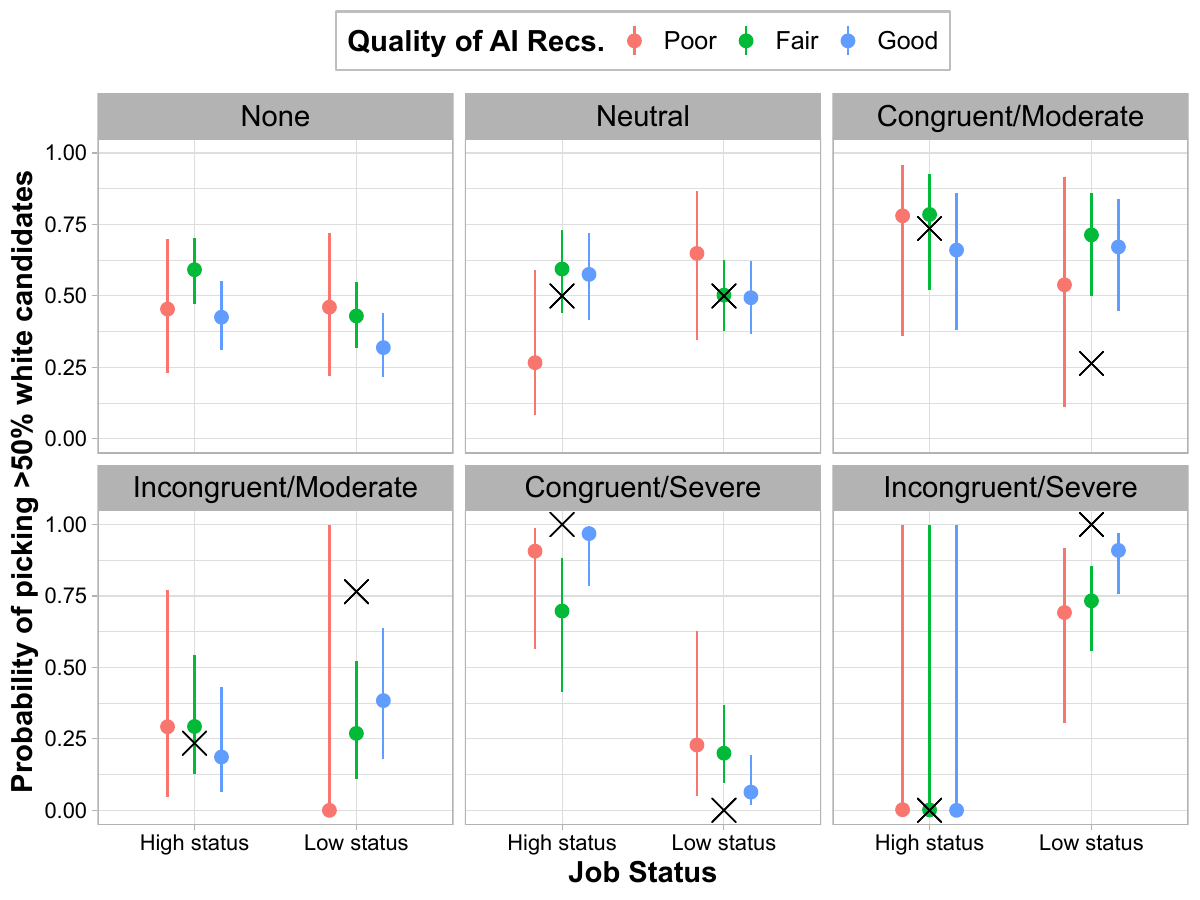}
\caption{Predicted probability of favoring white candidates in resume screening task split by response to AI recommendation quality. X marks the proportion of AI recommendations favoring white candidates.}
\label{fig:Quality_fig}
\end{figure*}

Figures \ref{fig:Importance_fig} and \ref{fig:Quality_fig} show the probability of preferring white candidates by participants' response to questions about AI recommendation importance and quality for all six bias magnitudes. 

\section{Task Instructions}
Welcome! We have developed some new artificial intelligence (AI) tools to help with hiring decisions. We are interested in what kinds of AI tools behave most like human decision makers and whether AI tools can improve the efficiency of making hiring decisions. 

\subsubsection{Decision-Making}
In this part of the task, you will be pretending to screen resumes in order to select candidates for interviews.

You will be given a job description as well as the profiles of several candidates, where some details have been redacted or altered in order to preserve the privacy of the candidates.

You might also be given information about whether an artificial intelligence (AI) model has predicted that a candidate should be interviewed or not.

If an AI model recommends a candidate, you will see a green check. If the candidate is not recommended, you will see a red cross mark. There may also be additional information about why the AI made a particular prediction that can be used to help you make your decisions. 

There will be four scenarios total, and you will have four minutes in each scenario to review all of the candidate profiles and make a decision about who should be interviewed.

When you are ready to begin, proceed to the next page.

\subsubsection{IAT}
In this part of the task, you will be asked to sort items into categories quickly. This is to test the speed and accuracy of your decision making, in order to determine whether artificial intelligence (AI) can increase people's efficiency when making decisions. 

You will place your left and right index fingers on the E and I keys. At the top of the screen you will see 2 categories. In the task, words and/or images appear in the middle of the screen. The table below shows the words and images you will see as well as the categories they belong to.

When the word/image belongs to the category on the left, press the E key as fast as you can. When it belongs to the category on the right, press the I key as fast as you can. If you make an error, a red X will appear. Correct errors by hitting the other key.

Please try to go as fast as you can while making as few errors as possible. 

\textsc{\{Table with attribute and target stimuli to be categorized\}}

\section{Debrief}
Thank you for participating in our study. Our research team would like to debrief you on the purpose of the study and what your role was in it. 

Our study aims to investigate how people make decisions when presented with generated information from artificial intelligence tools. Studies have shown individuals have implicit associations with concepts that can be related to stereotypes. In addition, researchers have also discovered that implicit associations also exist in artificial intelligence algorithms. 

In this study, we present you with information that is either congruent or incongruent with socially known implicit associations, for example, different gender associated with different occupations or different race associated with different pleasantness. The implicit association test is used to evaluate the implicit associations of individuals, which can help us investigate if individuals will choose information in a way that matches these associations. 

We would now like to ask for your feedback on the study. Did you feel comfortable with the instructions and the materials provided? Did you encounter any technical difficulties or experience confusion during the study? Please feel free to share any thoughts or concerns you may have. 

To wrap up, we want to remind you that your participation in this study was greatly appreciated and that your input was valuable to our research. The data we collected from you will be analyzed to inform how people interact with generated information from artificial intelligence and whether it relates to our implicit associations.

\section{IAT Interface}
\begin{figure}[!h]
    \centering
    \includegraphics[width=\linewidth]{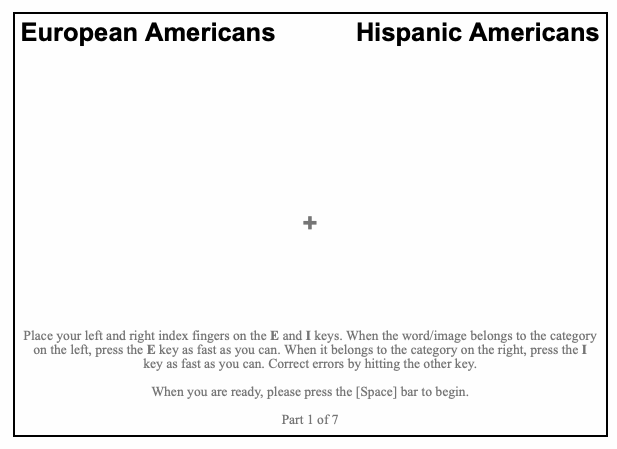}
    \caption{An example of the interface subjects saw when performing the IAT task. Instructions disappeared when the first item appeared at the fixation cross.}
    \label{fig:iat_interface}
\end{figure}

An example of the interface used to conduct IATs in Qualtrics with \texttt{iatgen} is in Figure \ref{fig:iat_interface}. In total subjects completed seven blocks in which the categories in the upper right and left corners changed between racial groups, high and low status words, or a combination of the two.

\section{AI Recommendation, Race, Task Order, Job
Status R Output}
\subsection{BLMM}
Table \ref{tab:GLMM_R} shows the output of the BLMM fit using \texttt{glmmTMB}, including fixed effects coefficients, AIC, log likelihood, and variance of random effects.

\input{Model1_GLMM.tex}

\subsection{ANOVA}
Table \ref{tab:R_anova} shows the output of the omnibus ANOVA conducted using the fit BLMM. Pairwise comparisons of significant omnibus effects are presented in the main text. 

\begin{table}[ht]
\centering
\begin{tabular}{lrrr}
  \hline
 & Chisq & Df & Pr($>$Chisq) \\ 
  \hline
(Intercept) & 3.41 & 1 & 0.0648 \\ 
  bias & 51.51 & 5 & 0.0000 \\ 
  Job\_type & 2.44 & 1 & 0.1185 \\ 
  Group\_recode & 1.22 & 2 & 0.5445 \\ 
  I\_recode & 0.01 & 1 & 0.9140 \\ 
  bias:Job\_type & 171.39 & 5 & 0.0000 \\ 
  bias:Group\_recode & 10.79 & 10 & 0.3739 \\ 
  Job\_type:Group\_recode & 0.29 & 2 & 0.8666 \\ 
  bias:I\_recode & 10.67 & 5 & 0.0582 \\ 
  Job\_type:I\_recode & 7.59 & 1 & 0.0059 \\ 
  Group\_recode:I\_recode & 5.41 & 2 & 0.0670 \\ 
  bias:Job\_type:Group\_recode & 7.39 & 10 & 0.6879 \\ 
  bias:Job\_type:I\_recode & 3.52 & 5 & 0.6200 \\ 
   \hline
\end{tabular}
\caption{\texttt{car} output for the ANOVA using the fit BLMM.}
\label{tab:R_anova}
\end{table}

\section{Exploratory Descriptive Statistics}
\begin{figure}[!h]
\centering
\includeinkscape[width=\linewidth]{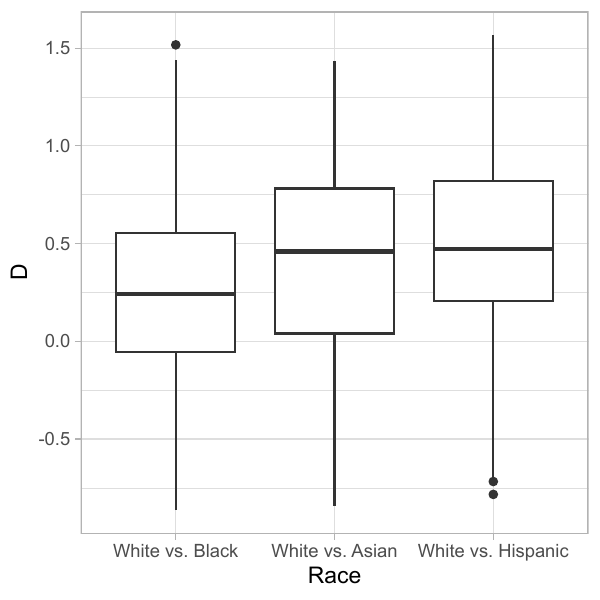}
\caption{Distribution of IAT scores for each \textit{Race} condition. Positive values indicate associations between white and high status; negative values indicate associations between non-white and high status.}
\label{fig:IAT_boxplot}
\end{figure}

\begin{figure}[!h]
\centering
\includeinkscape[width=\linewidth]{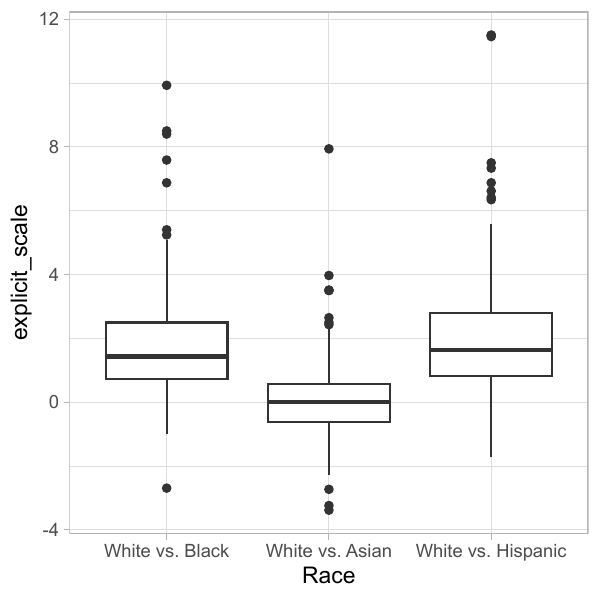}
\caption{Distribution of explicit belief  scores for each \textit{Race} condition. Positive values indicate stronger beliefs about white groups being high status; negative values indicate stronger beliefs about non-white groups being high status.}
\label{fig:explicit_boxplot}
\end{figure}

\begin{figure}[!h]
\centering
\includeinkscape[width=\linewidth]{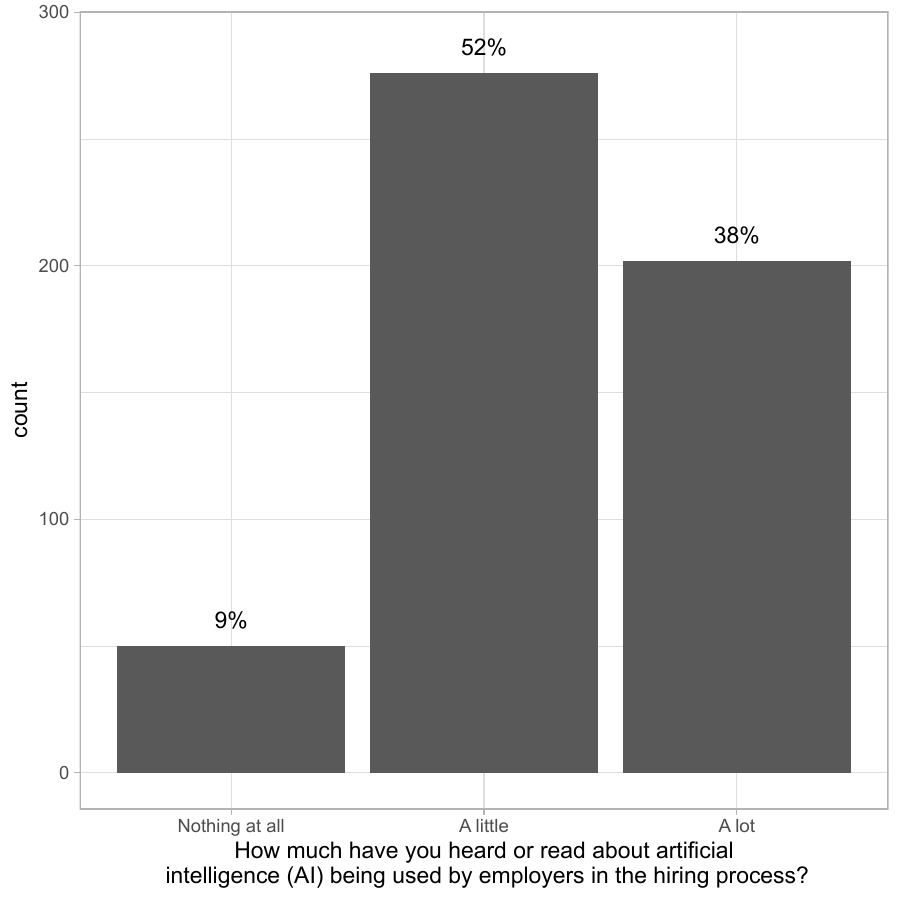}
\caption{Number of participant responses by each answer choice to the survey question about their previous AI experience.}
\label{fig:AIexp_barplot}
\end{figure}

\begin{figure}[!h]
\centering
\includeinkscape[width=\linewidth]{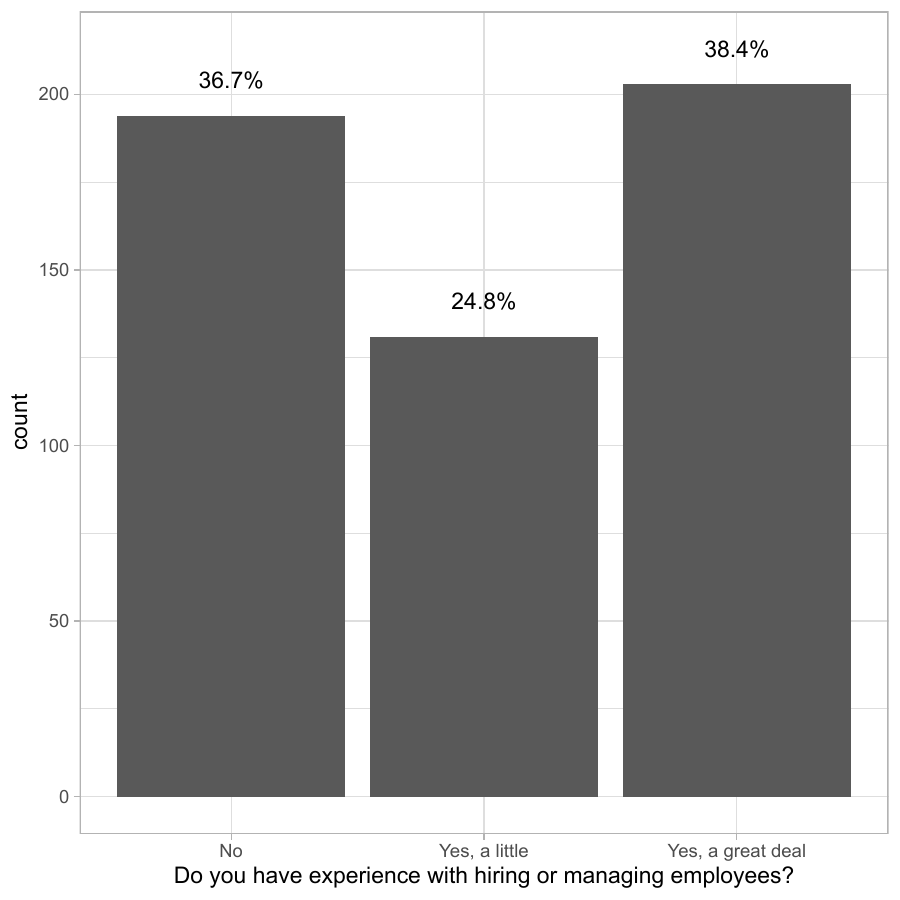}
\caption{Number of participant responses by each answer choice to the survey question about their previous hiring experience.}
\label{fig:HiringExp_barplot}
\end{figure}

\begin{figure}[!h]
\centering
\includeinkscape[width=\linewidth]{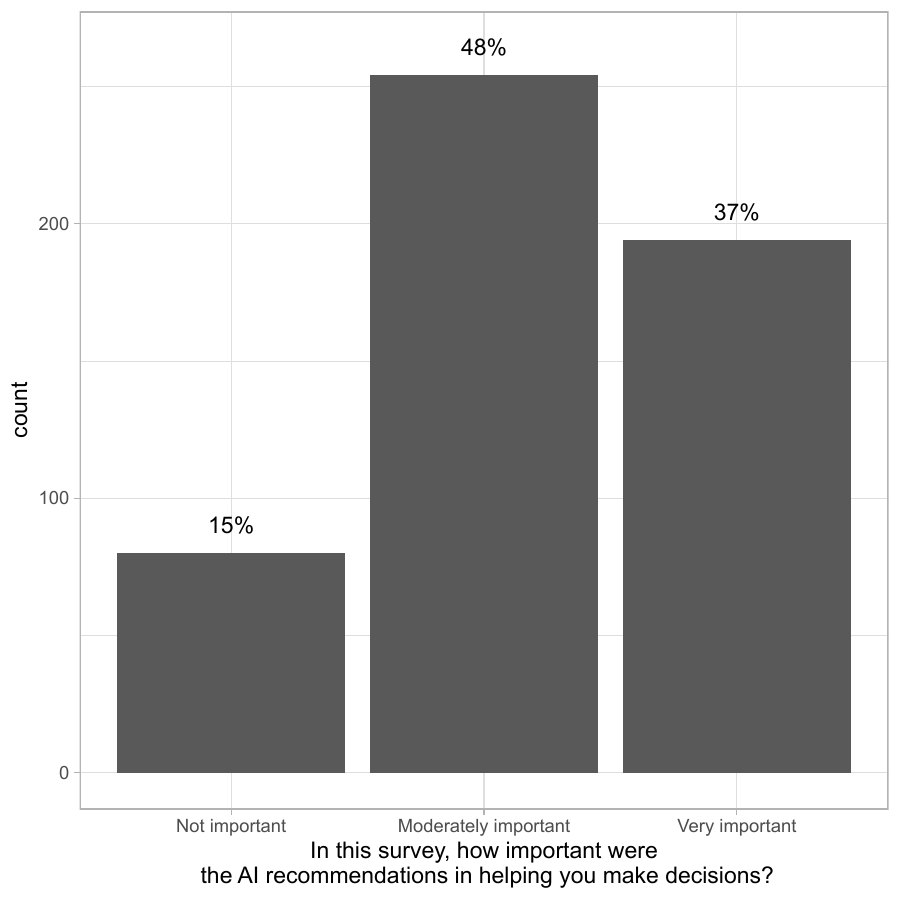}
\caption{Number of participant responses by each answer choice to the survey question about the importance of the AI hiring recommendations.}
\label{fig:Importance_barplot}
\end{figure}

\begin{figure}[!h]
\centering
\includeinkscape[width=\linewidth]{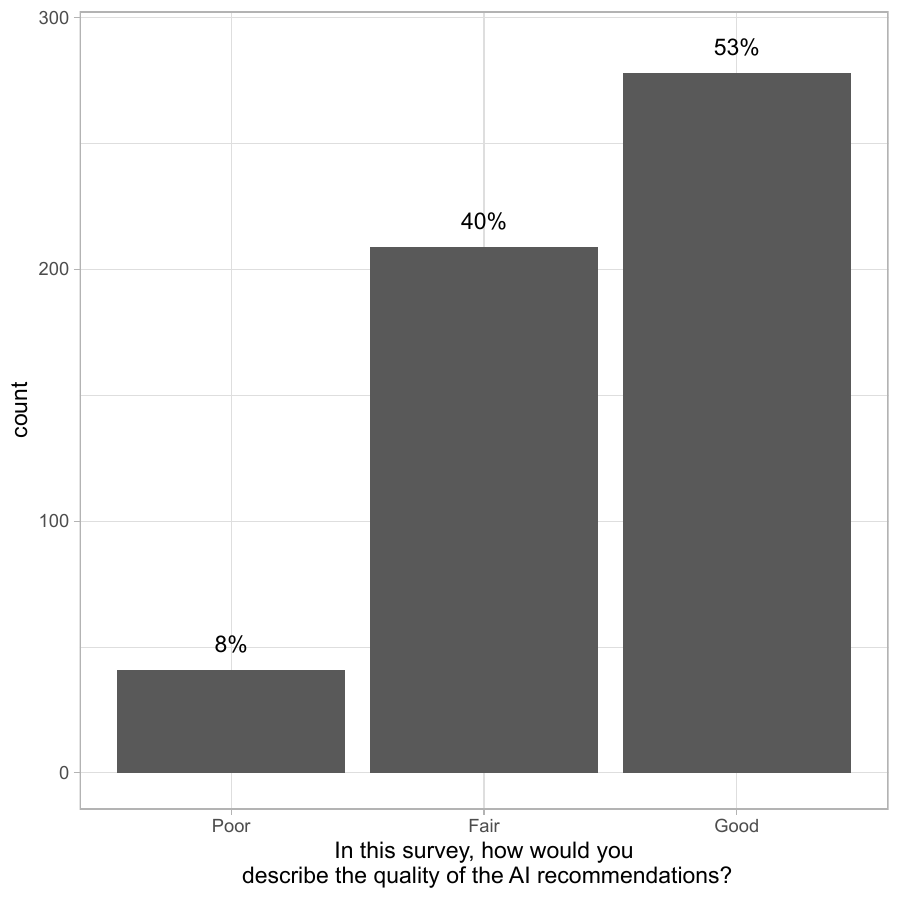}
\caption{Number of participant responses by each answer choice to the survey question about the quality of the AI hiring recommendations.}
\label{fig:Quality_barplot}
\end{figure}

\begin{figure}[!h]
    \centering
    \includegraphics[width=\linewidth]{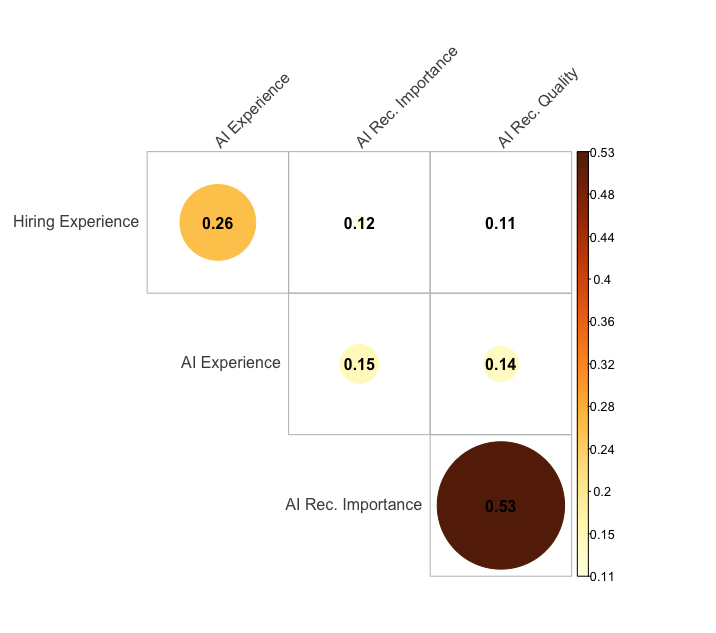}
    \caption{The strength of association between the categorical variables IAT score, explicit belief score, previous hiring and AI experience, and perceptions of AI recommendation quality and importance. Values greater between 0.2 and 0.6 indicate a moderate association.}
    \label{fig:exploratory_v}
\end{figure}

Figures \ref{fig:IAT_boxplot}-\ref{fig:Quality_barplot} show distributions of participant responses for the exploratory variables IAT score, explicit belief score, previous hiring and AI experience, and perceptions of AI recommendation quality and importance. Figure \ref{fig:exploratory_v} shows associations between the categorical variables using Cramer's V. There is a moderate association between people's previous experience with hiring and their familiarity with AI being used for hiring and between people's perceptions of AI recommendation quality and importance. 

\section{Exploratory Elastic Net} \label{sec:elastic}
\subsection{Approach}
As suggested by \citet{mcneish2015using}, we also used sparse regression techniques to confirm the exploratory results found using the backwards step-wise elimination procedure for the BLMM fitting. Specifically, we used the \texttt{grpnet} package to implement group penalized elastic net regression (GENR) to identify the most important terms in Equation 2 for predicting resume screening decision outcomes. This approach was selected for a number of reasons--first, the elastic net is a regularization penalty which balances the sparsity offered by the LASSO penalty and the inclusion of multiple correlated predictors by the Ridge penalty \citep{zou2005regularization}. The group penalty adds an additional constraint that all levels of a particular categorical factor should be included or excluded together \citep{helwig2025versatile}, which makes this approach particularly useful when very few predictors are continuous, as in this experiment \citep{huang2023lasso}. One limitation is that this approach is not able to account for random effects, so we apply the regularized regression to a a binomial logistic model (BLM) rather than a BLMM. This is warranted because the random effects for participant and job do not explain much variance in the data, as indicated by their elimination in the stepwise regression and a low intraclass correlation coefficient for the BLMM fit (ICC=.052).

We evaluated the GENR model based on its prediction capabilities and the importance of each predictor to the final model. For prediction, we partitioned the data so that 80\% of the original 1,526 trials was used to train the model and the remaining 20\% was used for evaluation. We also used 10-fold cross-validation over the training set to select the hyperparameters $\alpha$, which governs how much penalty weight is split between LASSO and Ridge penalties, and $\lambda$, which governs the strength of coefficient penalization. The best performing model used $\alpha=0.01$ and $\lambda=.515$. 

\subsection{Results}
Overall, the GENR model misclassified 33.784\% of trials in the test set, compared to 37.838\% using the GLMM in Equation 2. This indicates that a sparse model has better predictive performance than one that includes all possible variables and interactions. Table \ref{tab:elastic} shows the importance of each variable in the fit GENR. The interaction between \textit{AI Recommendation} and \textit{Job Status} explained the greatest amount of variance (29.57\%). AI recommendation quality, explicit score, AI recommendation importance, AI experience, and IAT score also had significant interactions with \textit{AI Recommendation} and \textit{Job Status} that explained at least 5\% of the variance in the model. This results confirm the findings presented in the main experiment that AI recommendation quality and importance are important to predicting people's decisions, and factors which did not necessarily emerge to be useful for causality (such as explicit and IAT scores) were still useful for prediction tasks.

\begin{table}[ht]
\small
\centering
\begin{tabular}{lr}
  \hline
\textbf{Variable} & \textbf{Importance} \\ 
  \hline
Job Status:AI Recommendation & 29.57 \\ 
  AI Recommendation:Job Status:Quality & 15.26 \\ 
  AI Recommendation:Job Status:Explicit Score & 10.08 \\ 
  AI Recommendation:Job Status:Importance & 9.63 \\ 
  AI Recommendation:Job Status:AI Experience & 8.08 \\ 
  AI Recommendation:Job Status:IAT Score & 6.12 \\ 
  AI Recommendation & 5.21 \\ 
  AI Recommendation:AI Experience & 2.62 \\ 
  AI Recommendation:Importance & 2.61 \\ 
  AI Recommendation:Quality & 1.89 \\ 
  Importance & 1.75 \\ 
  AI Recommendation:IAT Score & 1.46 \\ 
  AI Recommendation:Explicit Score & 1.36 \\ 
  Job Status:Task Order & 1.11 \\ 
  AI Recommendation:Task Order & 0.74 \\ 
  Explicit Score & 0.68 \\ 
  IAT Score & 0.43 \\ 
  AI Recommendation:Hiring Experience & 0.34 \\ 
  Job Status:IAT Score & 0.32 \\ 
  Race:Task Order & 0.24 \\ 
  AI Recommendation:Job Status:Hiring Experience & 0.14 \\ 
  AI Recommendation:Job Status:Task Order & 0.12 \\ 
  Job Status & 0.12 \\ 
  Race:Explicit Score & 0.11 \\ 
  Job Status:Importance & 0.10 \\ 
  Job Status:Explicit Score & 0.08 \\ 
  Quality & 0.07 \\ 
  Job Status:Quality & 0.06 \\ 
  Job Status:AI Experience & 0.03 \\ 
  Job Status:Hiring Experience & 0.03 \\ 
  Hiring Experience & -0.08 \\ 
  Task Order & -0.08 \\ 
  Race:IAT Score & -0.19 \\ 
   \hline
\end{tabular}
\caption{The importance of each predictor as a percentage of explained variance in the fit GENR.}
\label{tab:elastic}
\end{table}

\section{Exploratory Variable Importance}
\subsection{Approach}
As suggested by \citet{tonidandel2011relative}, we also calculated the importance of each variable to the overall fit of the model presented in Equation 2 to confirm the exploratory results found using the backwards step-wise elimination procedure for the BLMM fitting. Importance is measured by the total amount of deviance explained in the linear model by each variable either in isolation or in combination with other variables in the model, and we used the \texttt{tornado} package for calculating and plotting importance. Methods for calculating importance of mixed-effects models are still an open area of research \citep{diaz2015variable}, so we calculate importance based on a BLM rather than a BLMM, as was done in Section \ref{sec:elastic}.

\subsection{Results}
\begin{figure}[!ht]
\centering
\includeinkscape[width=\linewidth]{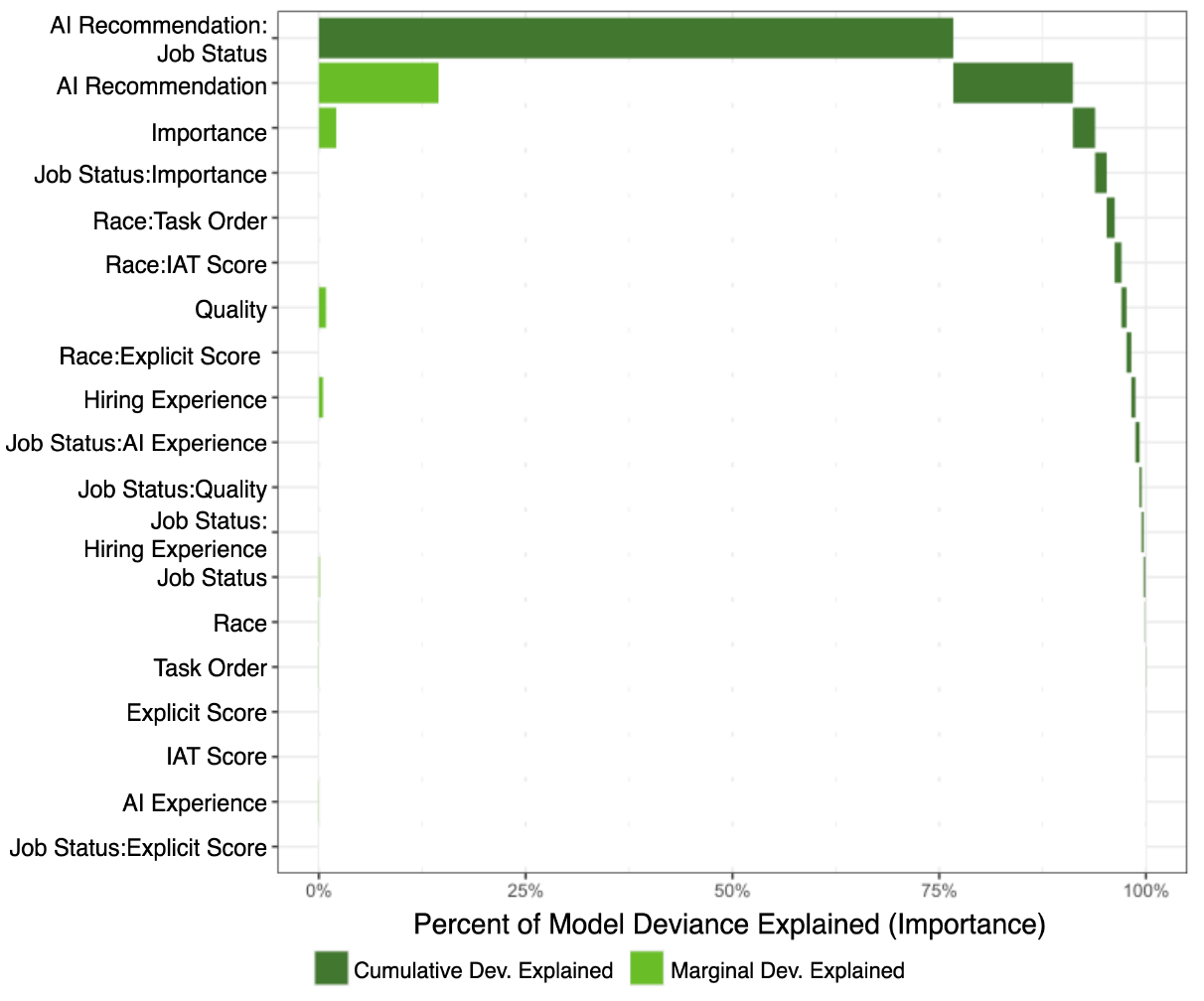}
\caption{The importance of each variable as a percentage of explained deviance.}
\label{fig:VarImportance_fig}
\end{figure}

Figure \ref{fig:VarImportance_fig} shows the results of the variable importance analysis. The interaction between \textit{Job Status} and \textit{AI Recommendation} explains 76.7\% of the total deviance while \textit{Job Status} and \textit{AI Recommendation} alone explain only 0.1\% and 14.5\% of deviance, respectively. Of the exploratory factors, AI recommendation importance explains 2.1\% of deviance, AI recommendation quality explains 0.9\% of deviance, and Hiring Experience explains 0.5\% of deviance. Importantly, we find that the two exploratory factors identified by the backwards stepwise elimination procedure, AI recommendation quality and importance, are also relevant to the model fit according to an alternative variable importance analysis.

%% file: Model1_GLMM.tex
\begin{table}
\begin{center}
\small
\begin{tabular}{l c}
\toprule
 & Model 1 \\
\midrule
(Intercept)                                  & $-0.19 \; (0.10)$       \\
biasNeutral                                  & $0.24 \; (0.15)$        \\
biasSim-Cong-New                             & $1.17 \; (0.24)^{***}$  \\
biasSim-Incong-New                           & $-0.97 \; (0.25)^{***}$ \\
biasExt-Cong                                 & $0.39 \; (0.23)$        \\
biasExt-Incong                               & $-0.09 \; (0.23)$       \\
Job\_type1                                   & $0.16 \; (0.10)$        \\
Group\_recode1                               & $0.06 \; (0.14)$        \\
Group\_recode2                               & $-0.16 \; (0.14)$       \\
I\_recode1                                   & $0.01 \; (0.10)$        \\
biasNeutral:Job\_type1                       & $-0.26 \; (0.15)$       \\
biasSim-Cong-New:Job\_type1                  & $-0.04 \; (0.23)$       \\
biasSim-Incong-New:Job\_type1                & $-0.10 \; (0.24)$       \\
biasExt-Cong:Job\_type1                      & $1.88 \; (0.24)^{***}$  \\
biasExt-Incong:Job\_type1                    & $-2.16 \; (0.24)^{***}$ \\
biasNeutral:Group\_recode1                   & $-0.06 \; (0.21)$       \\
biasSim-Cong-New:Group\_recode1              & $0.28 \; (0.33)$        \\
biasSim-Incong-New:Group\_recode1            & $-0.57 \; (0.36)$       \\
biasExt-Cong:Group\_recode1                  & $0.19 \; (0.33)$        \\
biasExt-Incong:Group\_recode1                & $-0.57 \; (0.35)$       \\
biasNeutral:Group\_recode2                   & $0.16 \; (0.21)$        \\
biasSim-Cong-New:Group\_recode2              & $0.38 \; (0.33)$        \\
biasSim-Incong-New:Group\_recode2            & $0.46 \; (0.34)$        \\
biasExt-Cong:Group\_recode2                  & $-0.06 \; (0.31)$       \\
biasExt-Incong:Group\_recode2                & $0.45 \; (0.31)$        \\
Job\_type1:Group\_recode1                    & $-0.05 \; (0.14)$       \\
Job\_type1:Group\_recode2                    & $0.07 \; (0.14)$        \\
biasNeutral:I\_recode1                       & $-0.20 \; (0.15)$       \\
biasSim-Cong-New:I\_recode1                  & $0.43 \; (0.24)$        \\
biasSim-Incong-New:I\_recode1                & $-0.16 \; (0.24)$       \\
biasExt-Cong:I\_recode1                      & $0.31 \; (0.23)$        \\
biasExt-Incong:I\_recode1                    & $-0.13 \; (0.22)$       \\
Job\_type1:I\_recode1                        & $0.28 \; (0.10)^{**}$   \\
Group\_recode1:I\_recode1                    & $0.11 \; (0.09)$        \\
Group\_recode2:I\_recode1                    & $-0.20 \; (0.09)^{*}$   \\
biasNeutral:Job\_type1:Group\_recode1        & $0.11 \; (0.21)$        \\
biasSim-Cong-New:Job\_type1:Group\_recode1   & $0.08 \; (0.33)$        \\
biasSim-Incong-New:Job\_type1:Group\_recode1 & $0.13 \; (0.36)$        \\
biasExt-Cong:Job\_type1:Group\_recode1       & $0.37 \; (0.33)$        \\
biasExt-Incong:Job\_type1:Group\_recode1     & $-0.16 \; (0.35)$       \\
biasNeutral:Job\_type1:Group\_recode2        & $0.13 \; (0.21)$        \\
biasSim-Cong-New:Job\_type1:Group\_recode2   & $-0.15 \; (0.33)$       \\
biasSim-Incong-New:Job\_type1:Group\_recode2 & $-0.55 \; (0.34)$       \\
biasExt-Cong:Job\_type1:Group\_recode2       & $-0.15 \; (0.31)$       \\
biasExt-Incong:Job\_type1:Group\_recode2     & $0.12 \; (0.31)$        \\
biasNeutral:Job\_type1:I\_recode1            & $-0.22 \; (0.15)$       \\
biasSim-Cong-New:Job\_type1:I\_recode1       & $-0.23 \; (0.24)$       \\
biasSim-Incong-New:Job\_type1:I\_recode1     & $-0.28 \; (0.24)$       \\
biasExt-Cong:Job\_type1:I\_recode1           & $-0.07 \; (0.23)$       \\
biasExt-Incong:Job\_type1:I\_recode1         & $-0.25 \; (0.22)$       \\
\midrule
AIC                                          & $1812.58$               \\
Log Likelihood                               & $-854.29$               \\
Num. obs.                                    & $1526$                  \\
Num. groups: PROLIFIC\_PID                   & $528$                   \\
Num. groups: JOB\_ID                         & $16$                    \\
Var: PROLIFIC\_PID (Intercept)               & $0.09$                  \\
Var: JOB\_ID (Intercept)                     & $0.00$                  \\
\bottomrule
\multicolumn{2}{l}{\scriptsize{$^{***}p<0.001$; $^{**}p<0.01$; $^{*}p<0.05$}}
\end{tabular}
\caption{\texttt{glmmTMB} output for BLMM model.}
\label{tab:GLMM_R}
\end{center}
\end{table}